\documentclass[11pt,a4paper,oneside]{article}
\pdfoutput=1
\usepackage{jheppub}
\usepackage{hyperref}
\usepackage{graphicx,xcolor}
\usepackage{sectsty,amssymb,amsmath}
\usepackage{booktabs}
\usepackage{slashed}
\usepackage[utf8]{inputenc} 
\usepackage{subcaption}
\usepackage{siunitx}
\usepackage{cleveref}
\usepackage{placeins}
\usepackage{float}
\usepackage{tikz}
\usepackage[normalem]{ulem}
\usepackage{mathtools}
\usepackage{soul}

\DeclareMathOperator{\Tr}{Tr}

\title{Hot Leptogenesis}

\author[a]{Michael~J.~Baker,}  \emailAdd{mjbaker@umass.edu}

\author[b]{Ansh Bhatnagar,} \emailAdd{ansh.bhatnagar@durham.ac.uk}

\author[b]{Djuna Croon,}
\emailAdd{djuna.l.croon@durham.ac.uk}

\author[b]{Jessica Turner}
\emailAdd{jessica.turner@durham.ac.uk }

\affiliation[a]{Department of Physics, University of Massachusetts Amherst, MA 01003, USA}

\affiliation[b]{Institute for Particle Physics Phenomenology, Department of Physics, Durham University, Durham DH1 3LE, U.K.}

\date{\today}

\preprint{IPPP/24/61}

\abstract{
We investigate a class of leptogenesis scenarios in which the sector containing the lightest right-handed neutrino establishes kinetic equilibrium at a temperature $T_{N_1} > T_\text{SM}$, where $T_\text{SM}$ is the temperature of the Standard Model sector. We study the reheating processes which realise this ``hot leptogenesis'' and the conditions under which kinetic and chemical equilibrium can be maintained. We derive and solve two sets of evolution equations, depending on the presence of chemical equilibrium within the hot sector, and numerically solve these for benchmark scenarios. We compare the viable parameter space of this model with standard leptogenesis scenarios with a thermal initial condition and find that hot leptogenesis resolves the neutrino and Higgs mass fine-tuning problems present in the standard scenario.
}

\begin{document}
\thispagestyle{empty}
\def\thefootnote{\fnsymbol{footnote}}
\setcounter{footnote}{1}

\setcounter{page}{0}
\maketitle
\vspace{-1cm}
\flushbottom

\def\thefootnote{\arabic{footnote}}
\setcounter{footnote}{0}
\newpage

\setstcolor{blue}

\section{Introduction}
\label{sec:intro}
The observed neutrino masses can elegantly be explained by a seesaw mechanism.  In the type-I seesaw mechanism \cite{Minkowski:1977sc, Yanagida:1979as, GellMann:1980vs, Mohapatra:1979ia}, at least two Majorana right-handed neutrinos (RHNs) are added to the Standard Model (SM):
\begin{align}
    \mathcal{L} \supset &\,
    i \bar{N}_i \slashed{\partial} N_i -\frac{1}{2} m_{N_i} \bar{N}_i^c N_i  - Y_{\alpha i} \bar{L}_\alpha \Tilde{\Phi} N_i + \rm{h.c.} \,,
\end{align}
where $i$ ($\alpha$) denotes RHN generational (lepton flavour) indices and are summed over,  the Yukawa matrix is given by $Y$ and the leptonic and Higgs doublets are given by $L^T = (\nu_L^T, l_L^T)$ and $\Phi$, respectively, with $\Tilde{\Phi} = i \sigma_2 \Phi $. Once the Higgs acquires a vacuum expectation value, the light neutrino masses are generated.

Besides providing a simple explanation for light neutrino masses, the type-I seesaw mechanism can also account for the observed baryon asymmetry via thermal leptogenesis~\cite{Fukugita:1986hr}.
In this scenario, a lepton asymmetry is produced through out-of-equilibrium and $CP$-violating decays of the RHNs~\cite{Covi:1996wh, Covi:1996fm, Pilaftsis:1997jf}. This lepton asymmetry is then converted into a baryon asymmetry via electroweak sphalerons.  Most leptogenesis calculations assume that the $N_1$ particles are in kinetic equilibrium with the Standard Model bath and inherit a thermal distribution with temperature $T_{\rm SM}$.
 For parameter choices which reproduce the observed neutrino masses, the heavier right-handed neutrinos, $N_2$ and $N_3$, are typically in both kinetic and chemical equilibrium with the SM thermal bath~\cite{Bernal:2017zvx}. 

While the standard leptogenesis mechanism can successfully generate neutrino masses consistent with data along with the observed baryon asymmetry, it typically leads to a tension~\cite{Clarke:2015gwa} between the Davidson-Ibarra bound~\cite{Davidson:2002qv} (a lower bound on the RHN masses, to achieve sufficient baryon asymmetry generation) and the Vissani bound~\cite{Vissani:1997ys} (an upper bound on the RHN masses, to avoid fine-tuning of the SM Higgs mass).  Furthermore, successful leptogenesis comes at the cost of an accidental cancellation between the tree and one-loop contributions to the light neutrino mass matrix \cite{Moffat:2018wke}. Resonant leptogenesis \cite{Pilaftsis:2003gt}, characterised by significantly lighter RHNs with a highly degenerate mass spectrum that enhances CP asymmetry during their decays, provides a way to lower the leptogenesis scale while addressing fine-tuning issues in both the Higgs and neutrino mass matrices.

\begin{figure}
\centering
\includegraphics[width=.9\textwidth]{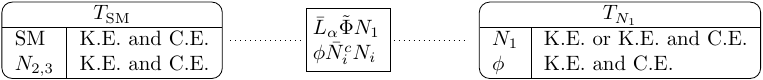}
\caption{Field content and the temperatures of the sectors in hot leptogenesis, along with whether they are in kinetic and chemical equilibrium (K.E. and C.E.) or only kinetic equilibrium (K.E.) around the time of $N_1$ decay.  The dominant coupling connecting the two sectors is taken to be the one responsible for $N_1$ decay.  The scalar field $\phi$, which keeps $N_1$ in kinetic equilibrium, may also mediate a coupling between $N_1$ and $N_{2,3}$ (and also between $N_1$ and the SM Higgs, not shown).  Particles of the scalar field $\phi$ may or may not be present at the time of $N_1$ decay, depending on whether $m_\phi$ is much greater than $T_{N_1}$ or not.}
\label{fig:hot-leptogenesis-sectors}
\end{figure}

In this work, we present an alternative solution to these tensions, where $N_1$ has a higher temperature, $T_{N_1}$, than the SM particles.  This leads to a larger number density of $N_1$ particles, which generate a larger baryon asymmetry after they decay.  This scenario has previously been considered in Ref.~\cite{Bernal:2017zvx}, where an enhancement of up to $\sim$ 50 times the standard leptogenesis baryon asymmetry can be obtained (the maximum exists because above some temperature the $N_1$ particles dominate the energy density of the universe).  Ref.~\cite{Bernal:2017zvx} requires a connection to thermal dark matter and concludes that resonant leptogenesis is still required to produce the observed baryon asymmetry.  In this work, we drop the connection to dark matter and show that non-resonant leptogenesis can produce the observed baryon asymmetry while remaining natural. Schematically, the scenario is depicted in \cref{fig:hot-leptogenesis-sectors}.  We will be interested in $T_\text{SM} < T_{N_1}$, and take the dominant coupling between the two sectors to be the one responsible for $N_1$ decay.  
Additional particles are needed to realise equilibrium within the hot sector. We quantitatively demonstrate that this can be realised by introducing a new scalar, $\phi$, which is
primarily responsible for mediating the self-interactions of $N_1$.  Within this setup, we consider two regimes.  In the first, $N_1$ is only in kinetic equilibrium during $N_1$ decay (so $N_1$ particles can exchange energy between themselves, but there are no number-changing processes that are sufficiently fast to realise chemical equilibrium).  In the second, $N_1$ is in both kinetic and chemical equilibrium with itself and $\phi$ (similar to the regime considered in Ref.~\cite{Bernal:2017zvx}).  That is, they are both in thermal equilibrium in the hot sector during $N_1$ decay.

In \cref{sec:inflationary-models} we first motivate this setup, showing that it can be a consequence of inflaton decays. In \cref{sec:model-of-hot-leptogenesis} we determine the regions of parameter space in our toy model where the two scenarios ($N_1$ in kinetic or kinetic and chemical equilibrium) are realised and discuss the cosmological constraints on the $\phi$ particle. In \cref{sec:evolution} we derive the relevant Boltzmann equations to track the evolution of the sectors and compute the resulting baryon asymmetry.  We present and discuss our results in \cref{sec:results} and conclude in \cref{sec:conclusions}.
\section{Hot Leptogenesis from Inflaton Decays}
\label{sec:inflationary-models}
While the mechanism of inflation is not yet determined, as a proof of principle we here discuss one plausible scenario.  The origin of two sectors with similar, but different, temperatures could be explained by an inflaton that couples with different strengths to the particles within each sector.  Assuming perturbative reheating, the reheating temperature in each sector is approximately,\footnote{Note that for efficient parametric resonance, the relation is more complicated. Parametric resonance is suppressed for inflaton decays to fermions due to Pauli blocking.}
\begin{equation}
  T_R 
  \approx 
  \sqrt{ \Gamma M_\text{Pl}}\,,
\end{equation}
where $\Gamma$ is the inflaton decay rate to particles within that sector and $M_\text{Pl} = 1.22 \times 10^{19}\,\rm{GeV} $ is the Planck mass. 
The decay rate of an inflaton $\sigma$, with mass $ m_\sigma$, to decay to particle species $i$, with mass $m_i$, is approximately
\begin{align}
    \Gamma_i 
    \approx 
    \frac{y^2 m_\sigma}{8 \pi}
    \,,
\end{align}
for $m_i \ll m_\sigma$ and where $y$ is the coupling of the inflaton to the particle. For an inflaton mass $m_\sigma \sim 10^{13}$ GeV, the reheating temperature is then $T_{R} \sim  y \times 10^{15}$ GeV, and the ratio of temperatures between the two sectors is
\begin{equation}
\label{eq:kappa_reh}
  \kappa 
  \equiv 
  \frac{T_{N_1}}{T_\text{SM}}
  \approx 
  \sqrt{\frac{\Gamma_{N_1}}{\Gamma_{\rm SM}}} 
  \approx 
  \frac{y_{N_1}}{y_{\rm SM}}\,,
\end{equation}
where $y_{N_1}$ ($y_{\rm SM}$) is the largest coupling of the inflaton to particles in the hot (SM) sector. As long as the two sectors cannot efficiently exchange energy, $y_\text{SM} < y_{N_1}$ will typically lead to $T_\text{SM} < T_{N_1}$.  While other factors can impact the precise value of $\kappa$, such as the spin of the daughter particles or other degrees of freedom with smaller couplings to $\sigma$, the fact that $y_{N_1}$ and $y_\text{SM}$ are not constrained by experiment mean that a wide range of values of $\kappa$ is plausible.

\begin{figure}
  \centering
  \begin{tabular}{ccc}
  \includegraphics[width=0.2\textwidth]{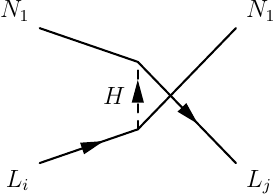}
  &
\includegraphics[width=0.2\textwidth]{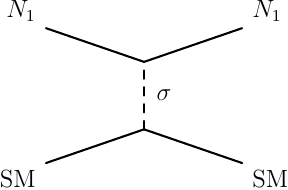}
  \\
  (a) & (b)
  \end{tabular}
  \caption{
  Feynman diagrams showing the (a) Higgs- and (b) inflaton-mediated processes which could thermalise the hot and SM sectors.  All Standard Model particles are denoted simply by SM.
  }
  \label{fig:inflaton-thermalisation}
\end{figure}
If there is only a weak coupling between the two sectors, they will not thermalise before the hot $N_1$ particles decay into particles in the  SM sector. A weak coupling means that leptogenesis will operate in the weak washout regime and ensures that scattering processes (such as those shown in \cref{fig:inflaton-thermalisation} (a)) are out-of-equilibrium before $N_1$ decays. In the case of strong washout, there are rapid interactions between the $N_1$ particles and the SM sectors which would cause the two sectors to thermalise.  It is also important that the inflaton itself does not thermalise the hot and SM sectors through the process shown in \cref{fig:inflaton-thermalisation} (b). To avoid this, we require the scattering rate between SM particles and $N_1$ particles to be slower than the Hubble rate.  Taking a simple Yukawa coupling between the inflaton and $N_1$,
\begin{align}
    \mathcal{L} \supset 
    \frac{1}{2}
    y_{N_1} \, \sigma 
    \bar{N}^c_1 N_1 \,,
\end{align} 
and assuming that the dominant inflaton-SM coupling is a universal Yukawa coupling to all SM fermions, 
\begin{align}
    \mathcal{L} \supset 
    y_\text{SM} \, \sigma 
    \sum_{f\in \text{SM}}
    \overline{\psi}_f \psi_f \,,
\end{align} 
we require
\begin{equation}
  \begin{split}
    \Gamma_{N_1\, \text{SM} \to N_1 \, \text{SM}} 
    = 
    \max(n_{N_1}, n_\text{SM}) \langle \sigma v \rangle 
    < 
    H \,,
  \end{split}
  \label{eq:noinfleq}
\end{equation}
where $n_{N_1}$ and $n_\text{SM}$ are the relevant number densities and $\langle \sigma v \rangle$ is the thermally averaged elastic cross-section between $N_1$ and the SM fermions via an inflaton mediator.  When the inflaton decays, the SM particles quickly reach thermal equilibrium so $n_\text{SM} = n_\text{SM}^\text{eq}(T_\text{SM})$.  We assume that some unspecified UV particles also allow $N_1$ to reach thermal equilibrium with itself around the reheating temperature, so that $n_{N_1} = 3\zeta(3)g_{N_1}T_{N_1}^3/4\pi^2$ where $g_{N_1} = 2$ is the number of degrees of freedom of $N_1$, but that these reactions rates fall below the Hubble rate before $N_1$ starts to decay (see \cref{sec:evolution}).  The cross-section for inflaton-mediated $N_1$--SM scatterings is
\begin{equation}
\label{eq:inflationcoupling}
      \sigma = \frac{ y_{\rm SM}^2 y_{N_1}^2}{4 \pi  s^2 \left(m_\sigma^2+s\right)} \left[s \left(2 m_\sigma^2-4 m_{N_{1}}^2+s\right) 
    -2 \left(2 m_{N_{1}}^2-m_\sigma^2\right) \left(m_\sigma^2+s\right) \log \left(\frac{m^2_\sigma}{m_\sigma^2+s}\right)\right]\,,
\end{equation}
where $s$ denotes the centre-of-mass energy, $m_{N_{1}}$ the mass of $N_1$ and we have assumed $m_f \ll m_{N_1}, m_\sigma$.
The thermal averaging for this process is discussed in \cref{App1}. For a universe consisting of two decoupled relativistic sectors with temperature ratio $\kappa$, the Hubble rate is given by
\begin{equation}
\label{eq:hubble}
  H = 
  \sqrt{\frac{8 \pi^3}{90}\left(g^*_{\rm SM} + g^*_{N_1} \kappa^4 \right)} 
  \frac{ T_{\rm SM}^2}{M_\text{Pl}}
  \,,
\end{equation}
where $g^*_{\rm SM}$ and $g^*_{N_1}$ denote the effective number of degrees of freedom in the Standard Model and hot sectors, respectively. The relative sizes of $g_\text{SM}^*$ and $g_{N_1}^* \kappa^4$ determine which sector dominates the energy density of the Universe.  For energies above the electroweak scale, with $g_\text{SM}^* = 106.75 + 4$ and assuming $g^*_{N_1} = 2$, the SM sector dominates the Universe's energy density for $\kappa \lesssim 2.7$, while the hot sector dominates for $2.7 \lesssim \kappa$. 
\begin{figure}
  \centering
  \includegraphics[width=.6\textwidth]{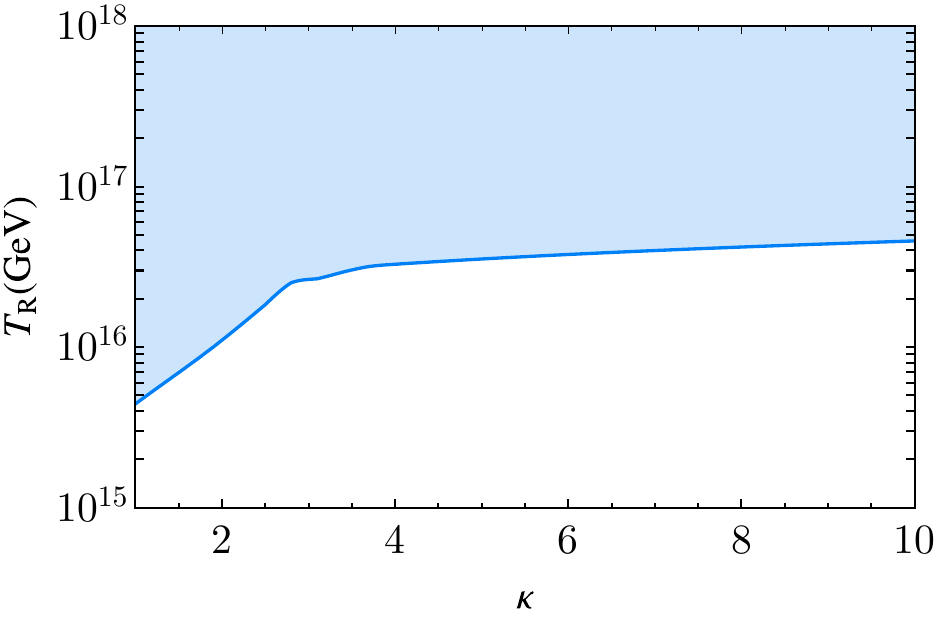}
  \caption{Upper bound on the reheating temperature in the hot sector from the requirement that inflaton-mediated elastic scattering does not realise kinetic equilibrium between the hot and SM sectors. In this plot we take $m_{N_1} = 10^7\,\text{GeV}$ and $m_\sigma = 10^{13}\,\text{GeV}$.}
  \label{fig:TRconstraint}
\end{figure}

Given \cref{eq:noinfleq,eq:inflationcoupling,eq:hubble}, we can determine the maximum value of $y_{\rm SM}y_{N_1}$ that ensures that the inflaton does not thermalise the two sectors.  For a given $\kappa$, this gives an upper bound on the reheating temperature of the hot sector.  For chaotic inflation, which fixes $m_\sigma \approx 10^{13}\,\text{GeV}$, and using $m_{N_1} = 10^7\,\text{GeV}$ as our benchmark value, the white region of \cref{fig:TRconstraint} shows the viable reheating temperatures of the hot sector as a function of $\kappa$. In the blue region, inflaton-mediated interactions will thermalise the two sectors, so that $T_\text{SM} \approx T_{N_1}$ and standard leptogenesis would proceed.  We find that for the two sectors to remain decoupled, we require $T_R \lesssim 10^{17}\,\rm GeV$, with a slight reduction when $\kappa \lesssim 3.8$ (where $n_{\rm SM} \lesssim n_{N_1}$). 
We will be interested in $N_1$ masses around $10^7\,\text{GeV}$, motivated by the Vissani bound limiting $m_{N_1} \lesssim 7.4 \times 10^7 \, \text{GeV}$ for the Higgs mass parameter $\mu^2$ to remain below $1\, \text{TeV}^2$ \cite{Vissani:1997ys}. Thus, the reheating temperature in the hot sector can be well above the right-handed neutrino masses. 

As mentioned above, the inflaton couplings to the different sectors are experimentally unconstrained.  As such, $\kappa$ can in principle take a wide range of values.  Some limiting scenarios often studied in the literature are:
\begin{enumerate}
  \item The inflaton decays exclusively to $N_1$~\cite{Lazarides:1990huy,Giudice:1999fb,Asaka:1999yd,Asaka:1999jb,Senoguz:2003hc,Hahn-Woernle:2008tsk}, corresponding to an initial $n_{\rm SM} = 0$ and $ \kappa \to \infty$. Such a scenario is typically studied in the context of non-thermal leptogenesis, where the assumption is that $T_R \ll m_{N_1}$ and that the $N_1$ decay happens immediately after the inflaton decay.  For example, Ref.~\cite{Giudice:1999fb} assumes that perturbative inflaton decay is kinematically forbidden, i.e., $m_\phi < 2 m_{N_1}$  so that the only relevant decay are through strong parametric resonance (overcoming Pauli-blocking of fermions). Ref.~\cite{Asaka:1999yd} studies perturbative inflaton decay into $N_1$, but with $100 T_R \lesssim m_{N_1}$ such that the $N_1$ particles are always out of kinetic and chemical equilibrium, making leptogenesis non-thermal. In this latter scenario, $N_1$ and the inflaton have similar masses, which is somewhat of a coincidence of scales. The inflaton decaying exclusively to $N_1$ was first studied away from the limit $T_R \ll m_{N_1}$ in Ref.~\cite{Hahn-Woernle:2008tsk}.  Without a self-interaction in the hot sector, the $N_1$ distribution after inflaton decay is non-thermal, and the Universe becomes radiation-dominated only after $N_1$ decay, complicating numerical analysis.  In our work, we will assume $m_{N_i} \ll T_R$ and the presence of an $N_1$ self-interaction, so the $N_1$ particles rapidly achieve a thermal distribution.
  \item The inflaton decay leads to $\kappa \approx 1$. When we take the case that $N_1$ are in kinetic and chemical equilibrium with themselves, see \cref{sec:chemicaleq}, our calculations with $\kappa = 1$ are comparable to the standard leptogenesis scenario with a thermal initial condition. 
  \item The inflaton decays only to the SM, corresponding to $\kappa \to 0$. This scenario corresponds to standard leptogenesis with a vanishing initial abundance of $N_1$.   Since we are interested in increasing the baryon asymmetry compared to standard leptogenesis by increasing the number density of $N_1$ particles in a sector that is hotter than the SM sector, we will not study $\kappa < 1$.
\end{enumerate}
In summary, we see that it is plausible for a simple model of inflation to lead to two decoupled sectors at a similar but different temperature, with a reheating temperature that is significantly above the right-handed neutrino masses. In what follows, we will take this as a starting point and we will study two different scenarios in the regime $1 \lesssim \kappa$. 
\section{A Model of Hot Leptogenesis}\label{sec:model-of-hot-leptogenesis}
While there are many possible realisations of the scenario we discuss, for concreteness we study a toy model consisting of the SM plus $N_2$ and $N_3$ at temperature $T_\text{SM}$ and a hot sector containing $N_1$ and a real scalar $\phi$ at temperature $T_{N_1}$.  The lightest right-handed neutrino, $N_1$, will decay to produce a lepton asymmetry which ultimately produces the baryon asymmetry, while $\phi$ will mediate interactions in the hot sector.  We will consider two cases: that around the time of $N_1$ decay either $N_1$ is only in kinetic equilibrium with itself, or $N_1$ is in both kinetic and chemical equilibrium with itself.  In this section, we find the regions of parameter space which exhibit these two cases. 

The relevant interaction Lagrangian terms in our toy model are,
\begin{align}
    \mathcal{L}
    \supset&\,
    - Y_{\alpha i} \bar{L}_\alpha \Tilde{\Phi} N_i + \rm{h.c.}
    \notag\\
    &\, - y_\phi^i \phi \bar{N}_i^c N_i - \frac{m^2_\phi}{2}\phi^2
    - \frac{\lambda_3 m_\phi}{3!} \phi^3
    - \frac{\lambda}{4!} \phi^4
    \,,
\end{align}
where $i \in \{1,2,3\}$ and $\alpha \in \{e,\mu,\tau\}$. The Yukawa matrix, $Y_{\alpha i}$, is parametrised using the Casas-Ibarra parameterisation \cite{Casas:2001sr}, $Y = \frac{1}{v} U \sqrt{{M}_\nu} R^T \sqrt{M_{N}}$, where $v = 174\,\text{GeV}$ is the vacuum expectation value of the Higgs, $U$ is the leptonic mixing matrix, $M_\nu$ ($M_{N}$) is the diagonal light (heavy) neutrino mass matrix and $R$ is a complex, orthogonal matrix. We discuss our specific choice of benchmark point for $Y_{\alpha i}$  in \cref{sec:results}.
We assume that the $\phi^3$ term is small enough that $\phi$ does not obtain a vacuum expectation value ($m_\phi^2 > 0 $ and $\lambda_3 < \sqrt{3\lambda}$). In principle, the scalar $\phi$ could generate the RHN masses, which is natural with diagonal couplings to the RHN mass eigenstates. However, we make the assumption that $\phi$ does not obtain a vev as we do not wish to restrict our attention to a particular mass generation mechanism and instead pursue a more general analysis of hot leptogenesis. Sizeable non-diagonal couplings between $\phi$ and the RHN generations could result in premature thermalisation of the two sectors, but we do not investigate this here. While in principle there are also cubic and quartic couplings with the SM Higgs, $\phi |\Phi|^2$ and $\phi^2 |\Phi|^2$, we assume that these are small enough to keep the two sectors out of thermal contact.  Although the inflaton could potentially play the role of $\phi$, we do not study this possibility and instead introduce a new scalar particle.
As noted in the introduction, standard leptogenesis leads to fine-tuning in the SM Higgs mass \cite{Vissani:1997ys,Clarke:2015gwa} and/or in the light neutrino masses \cite{Moffat:2018wke}.  The relevant expressions can be found in these references and the fine-tuning measures we use are given in \cref{app:fine-tuning}.

\subsection{Kinetic and Chemical Equilibria}
\label{sec:kinetic-and-chemical-equilibria}
We first consider the expected phase space distribution of $N_1$ in different regions of the parameter space of this model. 
 When $N_1$ is in kinetic and chemical equilibrium, it will have a Fermi-Dirac phase space distribution function, with zero chemical potential. 
When $N_1$ is only in kinetic equilibrium, its phase space distribution assumes the same form but will be normalised so that the number density of particles, $n_{N_1}$, is not fixed to the equilibrium number density,
\begin{align}
    f_{N_1}
    =
    \frac{n_{N_1}}{n^\text{eq}_{N_1}}
    f^\text{eq}_{N_1}
    \,.
\end{align}
Another possibility is that $N_1$ may have been in kinetic equilibrium at some point after inflaton decay but came out of kinetic equilibrium sometime before $N_1$ decay.  We do not analyse this case in detail, which would require the tracking of individual momentum modes, but we briefly discuss the expected applicability of our results to this scenario.  Finally, if the $N_1$ particles were never in kinetic equilibrium, their momentum would be spiked around half the inflaton mass.  We will not consider this case here.

There are a variety of processes to consider to determine which particles are in kinetic or chemical equilibria with themselves or each other.
The two sectors are necessarily coupled by the Lagrangian term responsible for $N_1$ decay, and potentially also by $\phi$ mediated processes.
The new scalar $\phi$ can keep $N_1$ in kinetic equilibrium with itself through $s$-, $t-$ and $u$-channel scattering processes, and if $\phi$ is not too much heavier than $N_1$ then number-changing interactions could also keep $N_1$ in chemical equilibrium with itself.
We now find the regions of parameter space where the following conditions hold:
\begin{enumerate}
    \item All elastic scattering processes between the hot and SM sectors are slower than the Hubble expansion rate.
    \item Elastic $N_1 N_1 \leftrightarrow N_1 N_1$ scattering processes are faster than the Hubble expansion rate.
    \item Number changing processes of both $N_1$ and $\phi$ are faster than the Hubble expansion rate. 
\end{enumerate}
Condition 1 ensures that the two sectors are decoupled, allowing each sector to maintain independent temperatures. Condition 2 ensures that $N_1$ is in kinetic equilibrium with itself, resulting in a Fermi-Dirac-shaped phase space distribution function. In our analyses, we will ensure that these two conditions always hold. If Condition 3 is satisfied, $N_1$ will be in both kinetic and chemical equilibrium, achieving an equilibrium number density. It is important to note that chemical equilibrium requires processes that can independently change the comoving number densities of $N_1$ and $\phi$; for example, the process $2 N_1 \leftrightarrow 2 \phi$ alone is not sufficient.

The Hubble rate in this model is given by
\begin{equation}
    \label{eq:hubble-toy-model}
  H = \frac{1}{\sqrt{3}M_\text{pl}}\sqrt{\frac{\pi^2}{30}g_\ast^\text{SM}(T_\text{SM})T_\text{SM}^4 + \rho_{N_1} + \rho_{N_2} + \rho_{N_3} + \rho_\phi}
  \,,
\end{equation}
where for $N_i$ and $\phi$
\begin{align} 
    \rho &= 
    \frac{n}{n^\text{eq}}
    \rho^\text{eq}
    =
    \frac{n}{n^\text{eq}}
    \frac{g}{2\pi^2}T^4 J_\pm\left(\frac{m}{T}\right)
    \,,
\end{align}
where
\begin{align}
    J_\pm(z) =&\, \int_0^\infty d\xi \frac{\xi^2\sqrt{\xi^2 + z^2}}{\exp [\sqrt{\xi^2 + z^2}] \pm 1} 
    \,,
\end{align}
with a plus sign for fermion, a negative sign for bosons and where $\xi = |\vec{p}|/T_{N_1}$.  For $N_2$ and $N_3$, which have a vanishing initial condition but approach their equilibrium energy densities throughout the evolution, for computational simplicity we approximate their contribution to the energy density as relativistic fermions in thermal equilibrium for $m_{N_{2,3}} < T_\text{SM}$ and we neglect their contribution otherwise. 
 We have checked that this approximation does not affect our final results. 

\subsubsection{Condition 1 -- Two Decoupled Sectors}
\label{sec:decoupled-sectors}
\begin{figure}
  \centering
  \begin{tabular}{ccc}
  \includegraphics[width=0.2\textwidth]{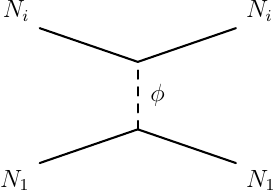}
  &
  \includegraphics[width=0.2\textwidth]{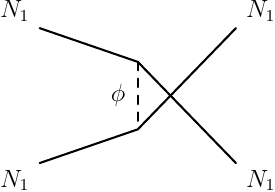}
  &
  \includegraphics[width=0.2\textwidth]{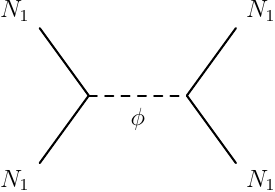}
  \\
  (a) & (b) & (c)
  \end{tabular}
  \caption{
  Feynman diagrams showing the processes which may put $N_1$ into kinetic equilibrium with itself ((a) with $i=1$, (b) and (c)) and with the SM bath ((a) with $i \in \{2,3\}$), which would set $T_\text{SM} = T_{N_1}$.
  }
  \label{fig:feyn-NNi-kinetic-eq}
\end{figure}
For the hot sector to remain thermally decoupled from the SM sector, we require that the decay rate of $N_1$ to SM particles is slower than the Hubble rate at $T_{N_1} \gtrsim m_{N_{1}}$. For $m_{N_1} \approx 10^7\,\text{GeV}$ this gives $Y_{\alpha 1} \lesssim 10^{-5}$.  This condition ensures leptogenesis proceeds in the weak washout regime and that the two sectors do not thermalise via inverse decays, $N_1 \text{SM} \to N_1 \text{SM}$ scatterings~\cite{Cline:1993bd,Garbrecht:2014kda} or processes like $N_1 \text{SM} \to \text{SM SM}$ before $N_1$ decays \cite{Bernal:2017zvx}.  This also ensures that other processes involving $Y_{\alpha 1}$, such as $N_1$ and $N_{2,3}$ thermalisation via the Higgs, are slower than the Hubble rate, due to the extra couplings and phase space suppression factors involved.

Beyond the direct coupling of $N_1$ with the SM plasma, it is possible that the $\phi$-mediated coupling between $N_1$ and the heavier $N_2$ and $N_3$ (which will be thermally produced in the SM sector) may thermalise the two sectors via the scattering process shown in \cref{fig:feyn-NNi-kinetic-eq} (a) with $i \in \{2,3\}$. 
We need to check that the interaction rate per $N_1$ particle and per $N_{2,3}$ particle is slower than the Hubble rate.  For $n_{N_1} \approx n^{\rm{eq}}_{N_1}$ we will have $n_{N_{2,3}} < n_{N_1}$ since $m_{N_1} < m_{N_{2,3}}$ and $T_\text{SM} < T_{N_1}$, so the rate per $N_1$ particle is slower than the rate per $N_{2,3}$ particle.  We therefore only need to check the rate per $N_{2,3}$ particle.  The two sectors will then not thermalise as long as
\begin{align}
 n_{N_1} \langle \sigma v\rangle_{N_1 N_{2,3} \to N_1 N_{2,3}} < H
    \,,
\end{align}
where the cross-section is given in \cref{eq:cs-N1NitoN1Ni}, the thermal averaging is given in \cref{eq:thermal-average-N1NitoN1Ni} and the Hubble rate is given in \cref{eq:hubble-toy-model}.  
Here, and for the remaining rate calculations, in this section we approximate  $n^{\rm{eq}}_{N_1} \approx n_{N_1}$. In principle, this could be modified when chemical equilibrium does not hold. This would lead to a proportional shift in the rates calculated here. Thus, our conclusions should be taken as a guide rather than precise statements for the kinetic equilibrium-only scenario.  However, it is a good approximation for the kinetic equilibrium-only cases we consider.

\begin{figure}
  \centering  \includegraphics[width=.95\textwidth]{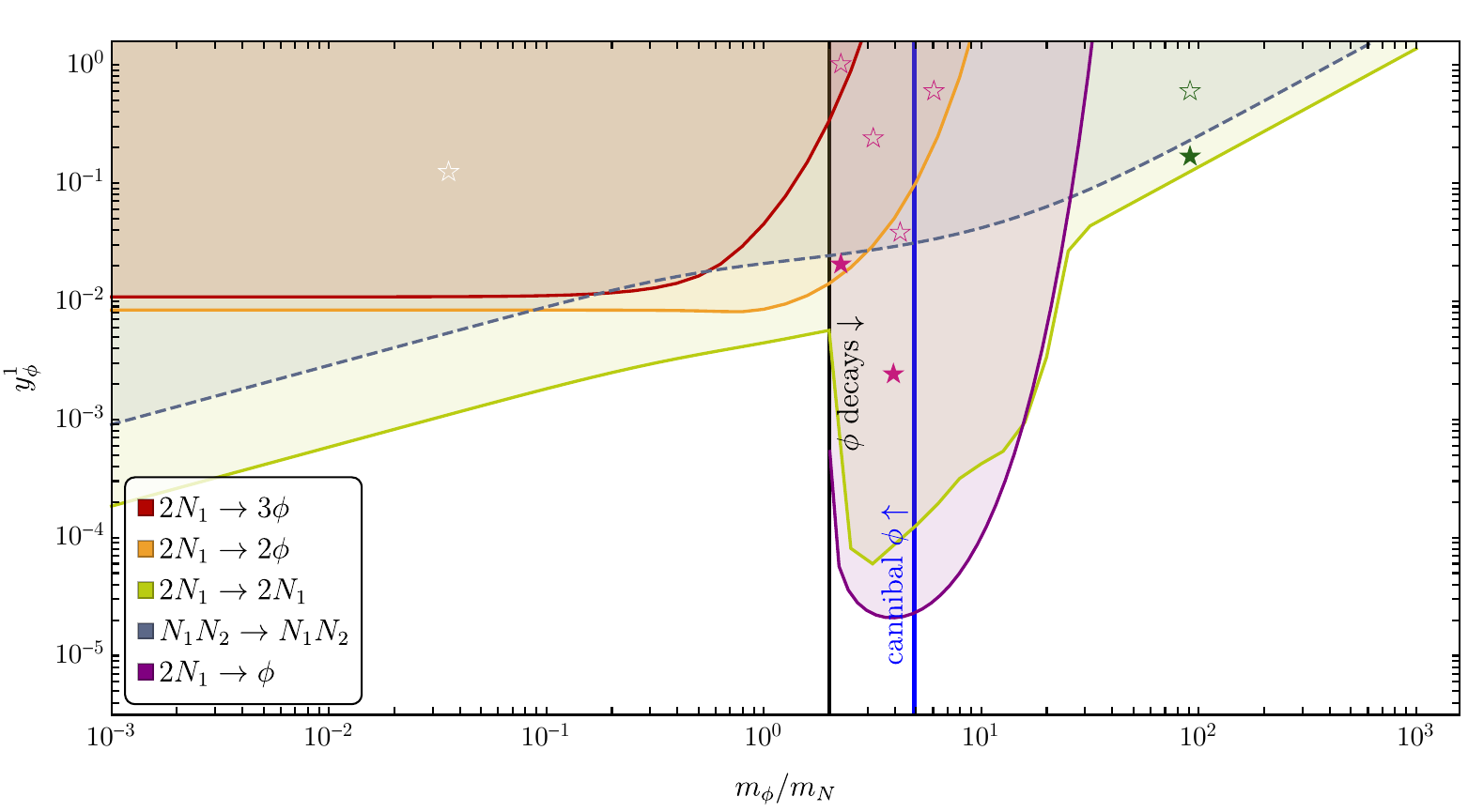}
  \caption{
  Minimal value of $y_\phi^1$ such that the various interaction rates are greater than Hubble around the time of decay, $T_{N_1} = m_{N_1} = 10^{7}\,\rm{GeV}$. We have assumed that Hubble is dominated by the SM as it is right after the decays and that $\lambda = 0.8$.  Assuming $y_\phi^1 = y_\phi^2$, the scattering process $N_1 N_2 \leftrightarrow N_1 N_2 $ will thermalise the SM and hot sectors in the blue region above the dashed blue contour.
  To the left of the black line the $\phi$ abundance will deplete with the $N_1$ abundance. To the left of the blue line (labelled ``cannibal $\phi$'') the cannibal process $2\phi \leftrightarrow 3 \phi$ is effective. The pink (green) stars indicate example points in the toy model parameter space where kinetic and chemical (only kinetic) equilibrium can be achieved, where open stars require $y_\phi^2, y_\phi^3 \ll y_\phi^1$.  The white star shows a point where the cosmology of $\phi$ would need to be carefully considered.
  }
  \label{fig:TEConstraints}
\end{figure}

We may expect the $N_1-\phi$ Yukawa coupling $y_\phi^1$ to be a similar order to $y_\phi^2$ and $y_\phi^3$ since the right-handed neutrino masses are all at a similar scale (although note that we do not discuss the origin of the right-handed neutrino masses here and do not assume that $\phi$ is a Majoron).  The blue region above the dashed blue contour in \cref{fig:TEConstraints} shows where the rate of this process is greater than the Hubble rate at the time of $N_1$ decay assuming $y_\phi^1 = y_\phi^2$.  That is, where condition 1 is not satisfied.    However, this bound can be relaxed, without affecting any other phenomenology, by taking $y_\phi^2 \ll y_\phi^1$.  Throughout this work, we consider the parameter space where the $N_1 N_{2,3} \leftrightarrow N_1 N_{2,3}$ scattering rate is less than the Hubble expansion rate to ensure the two sectors do not thermalise with each other.

There are also interactions between the $\phi$ and the Higgs that are induced at loop level via the coupling to the heavier RHN generations $N_{2,3}$, which typically have a larger Yukawa coupling to the Higgs than $N_1$. We computed the $\phi H H$ interaction rate at one-loop level and found an interaction rate much smaller than the Hubble rate $H \sim \frac{T^2}{M_p}$ for the scenarios we consider. Thus these interactions do not thermalise the two sectors.

\subsubsection{Condition 2 -- Kinetic Equilibrium within the Hot Sector}
\label{sec:kineticeq}

The processes shown in \cref{fig:feyn-NNi-kinetic-eq} (a) and (b) along with an extra $s$-channel process, \cref{fig:feyn-NNi-kinetic-eq} (c), will keep $N_1$ in kinetic equilibrium with itself if condition 2 is satisfied,
\begin{equation}
\label{eq:N_1-kin-eq-condition}
    n_{N_1}\langle \sigma v \rangle_{2N_1 \to 2N_1} > H \,,
\end{equation}
where the cross-section in given in \cref{eq:cs-2N1to2Ni} and the thermal averaging is given by \cref{eq:average}. The green region in \cref{fig:TEConstraints} shows where condition 2, or \cref{eq:N_1-kin-eq-condition}, is satisfied.  We see that there is a minimum value of $y_\phi^1$ for a given $m_\phi/m_{N_1}$.  When $2 m_{N_1} \lesssim m_\phi$ there is a resonant enhancement as the $\phi$ propagator in \cref{fig:feyn-NNi-kinetic-eq} (c) can go on-shell.

\subsubsection{Condition 3 -- Chemical Equilibrium Within the Hot Sector}
\label{sec:chemicaleq}

\begin{figure}
  \centering
  \begin{tabular}{cccc}
  \includegraphics[height=2.4cm]{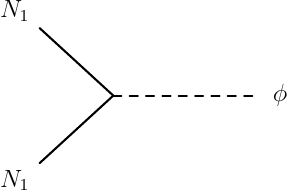}
  &
  \includegraphics[height=2.4cm]{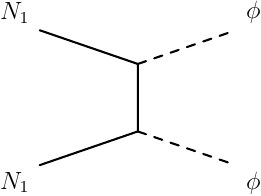}
  &
  \includegraphics[height=2.4cm]{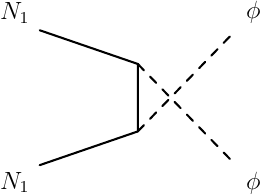}
  &
  \includegraphics[height=2.4cm]{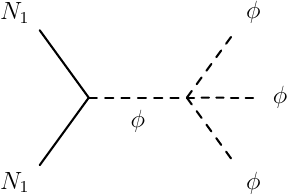}
  \\
  (a) & (b) & (c) & (d)
  \end{tabular}
  \caption{
  Feynman diagrams showing processes which could keep $N_1$ in chemical equilibrium with the hot sector.
  }
  \label{fig:feyn-NN-chemical-eq}
\end{figure}
If there are fast number-changing interactions, such as 
those shown in \cref{fig:feyn-NN-chemical-eq}, $N_1$ could also be in chemical equilibrium in the hot sector.
Chemical equilibrium requires processes which can increase (or decrease) both the comoving number densities of $N_1$ and $\phi$ at the same time.  This could, for instance, be a combination of $2 N_1 \leftrightarrow \phi$ and $2 N_1 \leftrightarrow 2 \phi$, or either of those along with $2\phi \leftrightarrow 3\phi$.  If these processes are faster than the Hubble rate we may assume that $n_{N_1}$ is simply given by $n_{N_1}^\text{eq}$, which simplifies the analysis. This is a version of the scenario considered in \cite{Bernal:2017zvx} (see, e.g., their eq.~2.19), where the hot sector was populated by the decay of dark matter. If only one of these processes is faster than the Hubble rate, neither $N_1$ nor $\phi$ can be assumed to be in chemical equilibrium and their abundances should be tracked dynamically.

$N_1$ and $\phi$ also both need to be in chemical equilibrium in the hot sector, so we technically need to check the rate per $N_1$ particle and the rate per $\phi$ particle.  However, since $n_\phi^\text{eq} < n_{N_1}^\text{eq}$ at $T_{N_1} = m_{N_1}$ for all $\phi$ masses, we only need to check the rate per $N_1$ particle. The purple region in \cref{fig:TEConstraints} shows where $\phi \leftrightarrow 2 N_1$ decays and inverse decays, \cref{fig:feyn-NN-chemical-eq} (a), occur faster than the Hubble rate,
\begin{align}
   \langle \Gamma_{\phi \to 2 N_1} \rangle 
   \frac{n_\phi^\text{eq}}{n_{N_1}^\text{eq}}
   = 
   \Gamma_{\phi \to 2N_1} 
   \frac{K_1(m_{N_1}/T_{N_1})}{K_2(m_{N_1}/T_{N_1})} 
   \frac{n_\phi^\text{eq}}{n_{N_1}^\text{eq}}
   > 
   H 
   \,,
\end{align}
where the $\phi \to 2 N_1$ decay rate is given in \cref{eq:gamma-2N1tophi} and $K_1$ and $K_2$ are modified Bessel functions of the second kind.  We see that in the range $2 m_{N_1} \lesssim m_\phi \lesssim 10 m_{N_1}$ this rate is faster than the Hubble rate if $10^{-4} \lesssim y_\phi^1$, and for larger $y_\phi^1$ this process can be relevant up to $m_\phi \approx 30 m_{N_1}$.

Next, we consider the process $2\phi \leftrightarrow 3 \phi$, which is independent of the coupling $y_\phi^1$.  Choosing a representative $\lambda = 0.8$ and $\lambda_3 = 0.57 \sqrt{3\lambda}$,\footnote{This value, which is not the result of spontaneous symmetry breaking~\cite{Hufnagel:2022aiz}, is chosen to maximise the cross-section.} we find that the region left of the blue vertical line satisfies
\begin{align}
  {n^2_\phi} \langle \sigma v^2 \rangle_{3\phi \to 2\phi} > H 
  \,,
\end{align}
where we take the thermally averaged cross-section from Ref.~\cite{Ghosh:2022asg}.  We see that, for these parameters, this process is faster than Hubble for $m_\phi < 5 m_{N_1}$.
In what follows, when we assume chemical equilibrium we will work in regions of parameter space where $2 N_1 \leftrightarrow 2 N_1$, $\phi \leftrightarrow 2 N_1$ and $2 \phi \leftrightarrow 3 \phi$ are all faster than Hubble.  However, we now briefly consider some other processes that could potentially be relevant. 

The orange region in \cref{fig:TEConstraints} shows where the $2 N_1 \leftrightarrow 2 \phi$ processes, \cref{fig:feyn-NN-chemical-eq} (b) and (c), are faster than the Hubble rate,
\begin{align}
  n_{N_1}\langle \sigma v \rangle_{2 N_1 \to 2 \phi} > H 
  \,,
\end{align}
where the cross-section is given in \cref{eq:cs-2N1to2phi} and the thermal averaging is given in \cref{eq:average}.  For $m_\phi \lesssim m_{N_1}$ this process is faster than Hubble if $10^{-2} \lesssim  y_\phi^1$, while for heavier $\phi$ particles a larger coupling is required.  

The process $2N_1 \leftrightarrow 3\phi$ can occur either via $t$-channel-like diagrams,  where the cross-section is proportional to $(y_\phi^1)^6$ or via an $s$-channel type diagram, \cref{fig:feyn-NN-chemical-eq} (d), whose cross-section is proportional $(y_\phi^1)^2\lambda^2$. Since we are mostly interested in the region $y_\phi^1 \ll \lambda \approx 1$, where the $t$-channel processes will be suppressed, we show in red in \cref{fig:TEConstraints} the region where the $s$-channel process is faster 
than Hubble,
\begin{equation}
  n_{N_1} \langle \sigma v \rangle_{2N_1 \to 3\phi} > H 
  \,.
\end{equation}
The cross-section for this process is given by \cref{eq:NNphi3} and the thermal averaging is done using \cref{eq:average}.  We see that this process is only faster than Hubble when $2N_1 \leftrightarrow 2\phi$ and $2 \phi \leftrightarrow 3 \phi$ are both faster than Hubble, so it does not create any new regions where chemical equilibrium for $N_1$ can be established (assuming similar values of $\lambda_3$ and $\lambda$ to those chosen here).

Another number-changing process that could be relevant is $2\phi \to 4\phi$. However, for an $n$-body massless particle, the phase space factor goes as
\begin{equation}
PS^{(n)} \sim \frac{1}{2(4\pi)^{2n-3} \Gamma(n)\Gamma(n-1)} \, .
\end{equation}
In comparison to the $2\phi \to 3\phi$ cross-section, there is a phase space suppression factor of $1/(192\pi^2)$. Therefore the ratio of the cross-sections goes as
\begin{equation}
    \frac{\sigma_{2\phi \to 4\phi}}{\sigma_{2\phi \to 3\phi}} 
    \sim 
    \frac{\lambda^2}{192 \pi^2\lambda_3^2 }  \, .
\end{equation}
This implies that the $2\phi \to 4\phi$ process is subdominant to $2\phi \to 3\phi$ as long as $\lambda \ll \sqrt{192} \pi \lambda_3 \approx 44 \lambda_3$. 

\subsection{Cosmology of the Scalar $\phi$}
We finish this section with a discussion of the cosmology of the new scalar $\phi$. If $m_\phi < 2m_{N_1}$, the hot sector will not deplete during $N_1$ decay and there will be a non-negligible abundance of $\phi$, which will freeze out relativistically. While the scalar $\phi$ is not stable, its four-body decay (via two off-shell $N_1$ particles) is suppressed by $Y_{\alpha i}^4$. 
If $\phi$ couples to all three RHNs, we find that $Y_{\alpha i}$ is typically large enough such that the decay can be expected to deplete the sector before BBN.  However, this will cause a large entropy dump which will wash out the generated asymmetry to some extent. If $\phi$ only couples to $N_1$, the smallness of $Y_{\alpha i}$ implies that $\phi$ is stable on a cosmological timescale. Because it freezes out relativistically, its large abundance will either give rise to a large contribution to $\Delta N_{\rm eff}$ -- the number of relativistic degrees of freedom at BBN -- or an overproduction of (dark) matter, depending on its mass. We conclude that in our current model $m_\phi$ needs to be larger than $2m_{N_1}$.\footnote{An alternative possibility is the existence of an additional portal coupling (for example a coupling to the SM Higgs) through which the $\phi$ abundance can deplete. In this scenario, the washout effect from this decay needs to be carefully considered.}
\subsection{Summary}
In summary, we see that there are many relevant processes and different possible regimes which can realise kinetic or kinetic and chemical equilibrium in the hot sector, even for our simple toy model.  To ensure the SM and hot sectors remain decoupled we require that leptogenesis occurs in the weak washout regime, effectively limiting the size of the Yukawa coupling.  From \cref{fig:TEConstraints} we see that, for $\lambda = 0.8$ and $\lambda_3 \sim \sqrt{3\lambda}$, we can assume that $N_1$ is in kinetic and chemical equilibrium with the hot sector around the time of $N_1$ decays if $2\, m_{N_1} < m_\phi < 5 \, m_{N_1}$, $10^{-4} \lesssim y_\phi^1$ and $y_\phi^{2,3} \lesssim 10^{-2}$. For $\lambda \ll 1$ chemical equilibrium could be established through $\phi \leftrightarrow 2N_1$ and $2 N_1 \leftrightarrow 2 \phi$ (which requires $y_\phi^1 \gtrsim 10^{-2}$ and typically require $y_\phi^{2,3} \ll y_\phi^1$).  When $30 \, m_{N_1} \lesssim m_\phi$ and $y_\phi^{2,3} \lesssim 10^{-2} \lesssim y_\phi^1$ we can assume that $N_1$ is only in kinetic equilibrium.  Both cases can also be realised for $m_\phi < 2m_{N_1}$, but in that case an entropy dump and/or washout would need to be carefully considered.  We indicate with pink, green and white stars the regions where these conditions hold (see caption of \cref{fig:TEConstraints} for details).

\section{Tracking the Evolution of the Hot and SM Sectors}
\label{sec:evolution}
In the simplest formulation, the leptogenesis kinetic equations operate in the one-flavoured regime, accounting for only a single flavour of charged lepton. This is a good approximation at very high temperatures ($T\gg 10^{12}$ GeV) when charged lepton Yukawa coupling processes are out of thermal equilibrium, resulting in a coherent superposition of the three flavour eigenstates. However, at lower temperatures ($10^{9}\,\text{GeV}\ll T  \ll 10^{12}\,\text{GeV}$), the interaction rates proportional to the tau Yukawa couplings come into thermal equilibrium and can cause decoherence, necessitating a description in terms of two flavour eigenstates. In our case, leptogenesis occurs at even lower temperatures ($T < 10^{9}$ GeV), where interactions mediated by the muon have equilibrated. In these regimes, a density matrix formalism \cite{Barbieri:1999ma, Abada:2006fw, DeSimone:2006nrs, Blanchet:2006ch, Blanchet:2011xq} provides a more comprehensive description than semiclassical Boltzmann equations, which do not include flavour oscillations in the lepton asymmetry. For this reason, we solve the density matrix equations which capture the time evolution of the RHN number densities in the hot and SM sectors, and the lepton asymmetry number density (which is promoted to a density matrix, $N_{\alpha \beta}$).
As discussed in \cref{sec:inflationary-models}, the Hubble expansion rate depends on the energy densities of both the hot and SM sectors. Since we track the energy density of both sectors, it will be convenient to evolve the density matrix equation as a function of the scale factor, $a$, which is then
\begin{align}
\label{eq:dNN1/da}
    aH \frac{d N_{N_1}}{d a} & =-\Gamma_{D_1}(z_{N_1})N_{N_1}+\Gamma_{D_1}(z_{\rm SM})N_{N_1}^{\mathrm{eq}} \,,
    \\
\label{eq:dNN2/da}
     aH\frac{d N_{N_2}}{d a} & =-\Gamma_{D_2}(z_{N_2})\left(N_{N_2}-N_{N_2}^{\mathrm{eq}}\right) \,,
    \\
\label{eq:dNN3/da}
    aH\frac{d N_{N_3}}{d a} & =-\Gamma_{D_3}(z_{N_3})\left(N_{N_3}-N_{N_3}^{\mathrm{eq}}\right) \,,
    \\
\label{eq:dNalphabeta/da}
    aH\frac{d N_{\alpha \beta}}{d a}&=\epsilon_{\alpha \beta}^{(1)} \left(\Gamma_{D_1}(z_{N_1})N_{N_1}-
    \Gamma_{D_1}(z_\text{SM})N^{\rm eq}_{N_1}\right)-\frac{1}{2} W_1\left\{P^{(1)}, n\right\}_{\alpha \beta} 
    \notag\\
    & +\sum_{i=2}^3\epsilon_{\alpha \beta}^{(i)} \Gamma_{D_i}(z_{N_i})\left(N_{N_i}-N_{N_i}^{\text {eq }}\right)-\frac{1}{2} W_i\left\{P^{(i)}, N\right\}_{\alpha \beta} 
    \notag\\
    & -\Lambda_\tau\left[\left(\begin{array}{lll}
    1 & 0 & 0 \\
    0 & 0 & 0 \\
    0 & 0 & 0
    \end{array}\right),\left[\left(\begin{array}{lll}
    1 & 0 & 0 \\
    0 & 0 & 0 \\
    0 & 0 & 0
    \end{array}\right), N\right]\right]_{\alpha \beta} 
    -\Lambda_\mu\left[\left(\begin{array}{lll}
    0 & 0 & 0 \\
    0 & 1 & 0 \\
    0 & 0 & 0
    \end{array}\right),\left[\left(\begin{array}{lll}
    0 & 0 & 0 \\
    0 & 1 & 0 \\
    0 & 0 & 0
    \end{array}\right), N\right]\right]_{\alpha \beta}\,,
\end{align}
where $i$ is a generation index, $\alpha, \beta$ are lepton  flavour indices, $N_{N_i}$ and $N_{\alpha \beta}$ are the comoving number density of $N_i$ and the $B-L$ asymmetry for lepton flavour indices $\alpha, \beta$, respectively, $\Gamma_{D_i} = \Gamma_{D_i}^0\langle m_{N_i}/E_{N_i} \rangle$ are the thermally averaged decay rates of $N_i$ where we assume Maxwell-Boltzmann statistics with \(\langle m_{N_i}/E_{N_i} \rangle = {K_1(z_{N_i})}/{K_2(z_{N_i})}\).
For the $N_1$ decay, we thermally average over the hot sector using the variable $z_{N_{1}}=m_{N_{1}}/T_{N_{1}}$ while for the $N_1$ inverse decays from the SM, and for \(N_2\) and \(N_3\), we thermally average over the Standard Model sector using the variables $z_{\rm SM} = m_{N_1}/T_{\rm SM}$ and $z_{N_{2,3}}=m_{N_{2,3}}/T_{\rm SM}$ respectively.
The equilibrium abundance of $N_i$ is denoted as ${N}_i^\text{eq}$ and the initial abundance of $N_1$ is $N_{N_{1}}\propto \kappa^3$. The initial abundance for $N_2$ and $N_3$ are assumed to be vanishing, which is an arbitrary choice for our computation as $N_2$ and $N_3$ are both in the strong washout regime so reach a thermal abundance before the $N_1$ particles decay. Thus, their initial abundance (whether thermal or vanishing) has an insignificant effect on the final results. The washout terms, which remove the lepton asymmetry produced by decays of $N_{1}$ in the hot sector, are denoted by $W_i$. We remind the reader that when the hot and visible sectors remain decoupled before $N_1$ decays, the washout is weak. The decay asymmetry (between RHNs decaying to leptonic and Higgs doublet versus the CP-conjugate process) generated by the decays of $N_i$ is given by the CP-asymmetry
matrix $\epsilon^{(i)}_{\alpha \beta}$ \cite{Covi:1996wh,Blanchet:2011xq,Abada:2006ea,DeSimone:2006nrs}. $\Lambda_\tau$ ($\Lambda_\mu$) denote the thermal widths of the tau (muon) charged leptons, which is obtained from the imaginary part of the self-energy correction to the lepton propagator in the plasma. Finally,  $P^{(i)}_{\alpha \beta} \equiv c_{i \alpha} c_{i \beta}^*$ where $c_{i \alpha}=Y_{\alpha i}/{\sqrt{\left(Y Y\right)_{i i}}}$ denote projection matrices which describe how a given flavour of lepton is washed out.
We note that this equation describes both the decays of $N_1$ from the hot sector into the SM and the possible inverse decays from the SM into the hot sector, as well as the SM washout processes. We note that while we include the evolution of the RHNs $N_{2}$ and $N_{3}$ for completeness, their contribution to the lepton asymmetry compared to $N_1$ is small. To compute the final lepton asymmetry one solves the coupled system for $z_{N_{1}}\gg 1$ and takes the trace of the $N_{\alpha \beta}$ matrix, $N^f_{B-L}=\Tr[N_{\alpha \beta}]$. Finally, to calculate the baryon asymmetry, we multiply $N^f_{B-L}$ by the sphaleron conversion factor and divide by the photon number density, to account for the change between the end of the leptogenesis era and recombination, $\eta_B = a_\text{sph} N^f_{B-L}/N_\gamma^\text{rec}$ where $ a_\text{sph} = 12/37$ \cite{DOnofrio:2014rug}. 

\subsection{$N_1$ in Kinetic and Chemical Equilibrium}

Having established the density matrix equations which determine the matter-antimatter asymmetry in our setup, we now consider how the temperatures of the two sectors evolve with time, assuming that kinetic and chemical equilibrium can be maintained within the hot sector while $N_1$ decays (pink stars in \cref{fig:TEConstraints}).  We first consider the SM temperature, $T_\text{SM}$, and then the hot sector temperature, $T_{N_1}$.

Before $N_1$ starts to decay around $T_{N_1} \sim m_{N_1}$, its comoving number density $N_{N_1}$ will remain constant, $T_{N_1}$ will drop as $a^{-1}$, and the two sectors will remain decoupled. 
When $N_1$ starts to decay, the hot sector transfers energy ($Q_{N_1}$) to the SM sector at a rate

\begin{align}
\label{eq:energy-transfer}
    \frac{dQ_{N_1}}{dt}
    =
    -\frac{dQ_\text{SM}}{dt}
    =
    - m_{N_1} V \Gamma_{D_1}^0\left( N_{N_1} -  N_{N_1}^{\mathrm{eq}}
    \right)
    \,,
\end{align}
where we normalise using the volume that contains one photon when we begin tracking the abundances, $V = 1/n^{eq}_\gamma(a=1)$.  Note that this is exact since the thermally averaged energy transfer rate is $\langle \Gamma_{D_1}^0 m_{N_1} E_{N_1}/E_{N_1}\rangle = m_{N_1}\Gamma_{D_1}^0$.  

As the SM sector is in thermodynamic equilibrium, we can apply the second law of thermodynamics to calculate the change in total entropy of the Standard Model,
\begin{align}
    \label{eq:second-law-sm}
    dS_\text{SM} &= \frac{d Q_\text{SM}}{T_\text{SM}}
    \,,
\end{align}
and use this to find the evolution of $T_\text{SM}$ with $a$.  First we write
\begin{align}
    \frac{d S_\text{SM}}{da}
    &=
    \frac{d (s_\text{SM} a^3 V)}{da}
    \,,
\end{align}
where $s_\text{SM}$ is the SM sector entropy density, which becomes
\begin{align}
    \label{eq:energy-entropy}
    \frac{1}{T_\text{SM}}\frac{d Q_\text{SM}}{da}
    &=
    a^3 V \frac{d s_\text{SM}}{d T_\text{SM}} \frac{d T_\text{SM}}{da} + 3 a^2 V s_\text{SM}
    \,,
\end{align}
 when we use \cref{eq:second-law-sm} and differentiate the right-hand side.  The rate of change of the SM sector entropy density $s_\text{SM}$ with respect to its temperature is
\begin{align}
    \label{eq:entropy-density-derivative}
  \frac{ds_\text{SM}}{dT_\text{SM}}=\frac{2\pi^2}{15}g_{\ast}(T_\text{SM}) T_\text{SM}^2
  +
  \frac{2\pi^2}{45}T_\text{SM}^3 \frac{d g_{\ast}(T_\text{SM})}{d T_\text{SM}}
  \,,
\end{align}
where numerically we neglect the second term which only has a small change due to $N_2$ and $N_3$.  Finally, using \cref{eq:energy-transfer,eq:energy-entropy,eq:entropy-density-derivative}, we find that
\begin{align}
    \label{eq:dTSM}
    \frac{d T_\text{SM}}{da}
    =
    \frac{m_{N_1}}{3 a^4 H s_\text{SM}}\Gamma_{D_1}^0\left( N_{N_1} -  N_{N_1}^{\mathrm{eq}}
    \right) 
    - 
    \frac{T_\text{SM}}{a}
    \,.
\end{align}

Next, we show how we determine the evolution of $T_{N_1}$.  When $N_1$ is in kinetic and chemical equilibrium, the solution to \cref{eq:dNN1/da} is the equilibrium comoving number density, which is
\begin{align} 
  \label{eq:nuNumber}
  N_{N_1}^\text{eq} =&\, a^3 n_{N_1}^\text{eq} V =\, a^3 V \frac{g_{N_1}}{2\pi^2}T_{N_1}^3 I_+\left(\frac{m_{N_1}}{T_{N_1}}\right) \,,
\end{align}
where $n_{N_1}$ is the (non-comoving) number density of $N_1$.  We account for the quantum statistics of $N_1$ using
\begin{align}
  I_+(z_{N_1}) =&\, \int_0^\infty d\xi \frac{\xi^2}{\exp [\sqrt{\xi^2 + z_{N_1}^2}] + 1}\,, 
\end{align}
where $\xi = |\vec{p}_{N_1}|/T_{N_1}$.  Thus, for a fixed $m_{N_1}$ there is then a one-to-one relationship between $N_{N_1} = N_{N_1}^\text{eq}$ and $T_{N_1}$, so solving \cref{eq:dNN1/da} tells us how $T_{N_1}$ evolves with $a$.
\Cref{eq:dNN1/da,eq:dNN2/da,eq:dNN3/da,eq:dNalphabeta/da,eq:dTSM,eq:nuNumber} then provide
a set of coupled differential equations which we solve using the numerical framework of {\sc ULYSSES} \cite{Granelli_2021,Granelli:2023vcm}.  Once we fix an initial $T_\text{SM}$, $\kappa = T_{N_1}/T_\text{SM}$ and the comoving number densities, we can use these equations to track $N_{N_i}$, $T_{N_1}$, $T_\text{SM}$ and $N_{\alpha\beta}$ as a function of $a$, which allows us to compute the final baryon asymmetry $\eta_B$. 
As the scalar $\phi$ remains in kinetic and chemical equilibrium with $N_1$, the hot sector depletes completely as $N_1$ decays.  We do not include the contribution from $\phi$ in the computation of the asymmetry, since for $2 m_{N_1} < m_\phi$ its abundance will be Boltzmann suppressed at the time of $N_1$ decay. 
 The $\phi$ population will both mildly increase the generated asymmetry by producing $N_1$'s as it decays, and mildly decrease it as it dumps entropy into the SM sector.
\subsection{$N_1$ in Kinetic Equilibrium Only}
We will now explore regions of the parameter space where number-changing interactions are slower than the Hubble rate, so the assumption of chemical equilibrium no longer holds (green stars in \cref{fig:TEConstraints}). In this case, the density matrix equations,  \cref{eq:dNN1/da,eq:dNN2/da,eq:dNN3/da,eq:dNalphabeta/da}, and the SM temperature derivative $dT_\text{SM}/da$, \cref{eq:dTSM}, remain the same as in the previous section. However, the evolution of the temperature of the hot sector, $T_{N_1}(a)$, is no longer given by \cref{eq:nuNumber}, as the lack of number-changing interactions leads to a departure from the equilibrium number density. 
The phase space distribution function for $N_1$ is $ f_{N_1} =     \left({n_{N_1}}/{n_{N_1}^\text{eq}} \right)f_{N_1}^\text{eq}$ where $f_{N_1}^\text{eq}$ is the Fermi-Dirac distribution.  Approximating the Fermi-Dirac distribution with a Maxwell-Boltzmann distribution, the energy density $\rho$ and pressure $p$ of $N_1$ become
\begin{align}
    \label{eq:kin-only-energy-density}
    \rho_{N_1} &= 
    \left(\frac{n_{N_1}}{n_{N_1}^\text{eq}} \right)
    \rho_{N_1}^\text{eq}
    \,,
    \\
    \label{eq:kin-only-pressure}
    p_{N_1} &= 
    \left(\frac{n_{N_1}}{n_{N_1}^\text{eq}} \right)
    p_{N_1}^\text{eq}
    \,.
\end{align}

We can now calculate the evolution of $T_{N_{1}}$ using the second law of thermodynamics and comoving energy conservation. First, we use the second law of thermodynamics,
\begin{align}
    dS_{N_1} = \frac{dQ_{N_1}}{T_{N_1}}
    \,,
\end{align}
to equate
\begin{align}
    \frac{dS_{N_1}}{da}
    &=
    \frac{ds_{N_1} a^3 V}{da}
    \\
    &=
    \frac{d}{da} \left( \frac{\rho_{N_1} + p_{N_1}}{T_{N_1}} a^3 V \right)
    \\
    \label{eq:dSN1-da}
    &=
    \frac{a^3V}{T_{N_{1}}}
    \left(
    \frac{d\rho_{N_1}}{da}
    +
    \frac{dp_{N_1}}{da} 
    - 
    s_{N_1} \frac{dT_{N_{1}}}{da} 
    +
    3\frac{s_{N_1} T_{N_{1}}}{a}
    \right)
    \,,
\end{align}
with
\begin{align}
    \label{eq:dQN1-da}
    \frac{1}{T_{N_1}} \frac{dQ_{N_1}}{da}
    =
    -\frac{m_{N_1}V}{T_{N_1}} \left(\Gamma_{D_1}^0 N_{N_1} -  \Gamma_{D_1}^0N_{N_1}^{\mathrm{eq}}
    \right)
    \,.
\end{align}
We see that we now require expressions for the rate of change of the energy density of $N_1$, $d \rho_{N_1} / da$, and for the rate of change of the pressure of $N_1$, $d p_{N_1} / da$.  

To find an expression for $d \rho_{N_1} / da$ we can use the conservation of total comoving energy density, 
\begin{align}
    a\frac{d\rho_\text{tot}}{da} + 3(\rho_\text{tot} + p_\text{tot}) = 0
    \,,
\end{align}
where $\rho_\text{tot}$ and $p_\text{tot}$ are the total energy density and pressure, respectively. We see that the derivatives of the energy densities in the two sectors are related by
\begin{equation}
a\frac{d\rho_{N_1}}{da} = - 3(\rho_\text{tot}+p_\text{tot}) - a\frac{d\rho_\text{SM}}{da} \, .
\label{eq:energycon}
\end{equation}
For the rate of change of the pressure, $d p_{N_1} / da$, we differentiate \cref{eq:kin-only-pressure} with respect to $a$ to write $d p_{N_1} / da$ in terms of $d T_{N_1} / da$ and $d n_{N_1} / da$,
\begin{align}
    \frac{dp_{N_1}}{da} 
    &= 
    \frac{n_{N_1}}{n_{N_1}^\text{eq}} \frac{dp_{N_1}^\text{eq}}{da} 
    + 
    \frac{dn_{N_1}}{da} \frac{p_{N_1}^\text{eq}}{n_{N_1}^\text{eq}} 
    - 
    \frac{p_{N_1}^\text{eq}}{n_{N_1}^\text{eq}} \frac{dn_{N_1}^\text{eq}}{da} 
    \\
\label{eq:pressurederiv}
    &= 
    \left(\frac{n_{N_1}}{n_{N_1}^\text{eq}}\frac{dp_{N_1}^\text{eq}}{dT_{N_{1}}} 
    - 
    \frac{p_{N_1}^\text{eq}}{n_{N_1}^\text{eq}} \frac{dn_{N_1}^\text{eq}}{dT_{N_{1}}} \right)\frac{dT_{N_{1}}}{da} 
    + 
    \frac{dn_{N_1}}{da} \frac{p_{N_1}^\text{eq}}{n_{N_1}^\text{eq}} \,.
\end{align}
Together, \cref{eq:dSN1-da,eq:dQN1-da,eq:energycon,eq:pressurederiv} give an expression for $dT_{N_1}/da$ in terms of quantities we can compute or track.
This, in combination with the density matrix equations, \cref{eq:dNN1/da,eq:dNN2/da,eq:dNN3/da,eq:dNalphabeta/da}, and the $T_\text{SM}$ evolution equation in the previous section, \cref{eq:dTSM}, can be solved with the appropriate initial conditions to find the resulting baryon asymmetry.  The initial conditions we take for our benchmark point are that $N_1$ has an equilibrium abundance while $N_2$ and $N_3$ have vanishing initial abundance.  This is motivated by the possibility that $\phi$ (or another particle) mediated sufficiently fast number changing rates for $N_1$ at higher temperatures, but that it no longer can at $T_{N_1} \sim m_{N_1}$. In the case where $N_1$ is only in kinetic equilibrium with itself around its decay, the $\phi$ abundance is heavily suppressed and does not contribute to the asymmetry generation (beyond maintaining kinetic equilibrium in the hot sector).
\section{Results}
\label{sec:results}
Standard non-resonant leptogenesis, which can produce a baryon asymmetry consistent with observations using RHN masses around $10^6$--$10^7$ GeV, requires fine-tuning of the light neutrino masses \cite{Moffat:2018wke} and the SM Higgs mass~\cite{Vissani:1997ys,Clarke:2015gwa}. The neutrino mass fine-tuning occurs because the tree and one-loop contributions to the light neutrino mass matrix have opposing signs~\cite{Lopez-Pavon:2012yda} and standard leptogenesis requires them to be separately large and then largely cancel to give an overall small neutrino mass consistent with observation. Fine-tuned solutions are favoured because the specific structure of the $R$-matrix reduces effective Yukawa couplings, which in turn decreases washout effects and leads to successful leptogenesis, while also enabling this critical cancellation between the tree-level and one-loop contributions, keeping the light neutrino masses within experimental bounds. The SM Higgs mass is also fine-tuned as the RHNs contribute at the one-loop level, with lower RHN masses giving a smaller loop level contribution but also reducing the amount of baryon asymmetry produced. 

\begin{table}
    \centering
    \begin{tabular}{cccccccccc}
    \toprule
        Benchmark & $S_1$ & $S_2$ & $S_3$ & $\overline{S_1}$ & $\overline{S_2}$ & $\overline{S_3}$ & $S_4$ & $\overline{S_4}$ & Our Benchmark\\
        \hline
        $\Delta_\nu\,[\%]$ & 0.2 & 0.3 & 0.2 & 0.6 & 0.7 & 0.4 & 0.2 & 0.5 & 855\\ 
        $\Delta_H\,[\%]$ & 0.08 & 0.02 & 0.004 & 0.3 & 0.06 & 0.1 & 0.001 & 0.007 & 10.4\\
        \bottomrule
    \end{tabular}
    \caption{Degree of fine-tuning for the best-fit points found in Ref.~\cite{Moffat:2018wke} and for our benchmark point (see \cref{tab:params}), using the fine-tuning measures given in \cref{app:fine-tuning}.  Smaller numbers indicate a larger degree of fine-tuning, with some degree of fine-tuning for numbers smaller than $\sim 10\%$.
    }
    \label{tab:fine-tuning}
\end{table}

Using the measures defined in \cref{app:fine-tuning}, the degree of fine-tuning for both the light neutrino mass and the SM Higgs required for successful standard leptogenesis with $m_{N_1} \sim 10^{6.5}$ GeV is given in \cref{tab:fine-tuning}, based on benchmarks from Ref.~\cite{Moffat:2018wke}. These benchmarks, $S_i$ ($\overline{S_i}$) for $i=1,2,3,4$, correspond to normal (inverted) ordering with the PMNS matrix parameters set to their best-fit values based on global data \cite{Esteban:2016qun}. The remaining CI parameters are fixed to ensure a viable baryon asymmetry. From \cref{tab:fine-tuning}, we see that the fine-tuning is worse than 1\,\% for both the light neutrino mass and the SM Higgs. We present these benchmarks and their associated tuning measures to evaluate how effectively hot leptogenesis can reduce fine-tuning while still producing the observed baryon asymmetry.
\begin{table}[t!]
    \centering
    \begin{tabular}{cc|c}
    \toprule
        Parameter & Unit & Benchmark point \\ \hline
        $\delta~$                       & $\left[^\circ\right]$      & 270 \\
        $\alpha_{21}~$                  & $\left[^\circ\right]$      & 50\\
        $\alpha_{31}~$                  & $\left[^\circ\right]$      & 120\\
        $\theta_{23}~$                  & $\left[^\circ\right]$      & 49.1\\
        $\theta_{12}~$                  & $\left[^\circ\right]$      & 33.41\\
        $\theta_{13}~$                  & $\left[^\circ\right]$      & 8.54\\ \hline
        $ x_1~$                       & $\left[^\circ\right]$      & 7\\
        $ y_1~$                       & $\left[^\circ\right]$      & 15\\
        $ x_2~$                           & $\left[^\circ\right]$      & 1\\
        $ y_2~$                       & $\left[^\circ\right]$      & 2\\
        $ x_3~$                           & $\left[^\circ\right]$      & 4\\
        $ y_3~$                       & $\left[^\circ\right]$      & 3\\ \hline
        $m_{1}$ & $\left[\mathrm{eV}\right]$ & 0\\
        $\log_{10}\left(m_{N_1}/[{\rm GeV}]\right)$    &    [1]                 & 7\\
        $\log_{10}\left(m_{N_2}/[{\rm GeV}]\right)$    & [1]                   & 7.006\\
        $\log_{10}\left(m_{N_3}/[{\rm GeV}]\right)$    &   [1]                  & 7.4\\ \hline
        $\kappa$    & [$1$]                    & 10\\
            \bottomrule
    \end{tabular}
    \caption{Input parameters in the Casas-Ibarra parametrisation for our benchmark point, see text for details.}
    \label{tab:params}
\end{table}

Here we explore two distinct scenarios of hot leptogenesis which alleviate this fine-tuning: one in which $N_1$ is only in kinetic equilibrium with itself before decay and one in which chemical equilibrium is also established. The parameters that determine whether chemical and kinetic equilibrium are established are different to those that determine the baryon asymmetry (except that they determine the manner of evolution of our model -- these scenarios evolve under a different set of evolution equations as described in the previous section).  The parameters related to the baryon asymmetry are given in \cref{tab:params}.  We will focus on a benchmark scenario for the neutrino parameters, where we use the Casas-Ibarra parametrisation to construct the Yukawa matrix $Y$~\cite{Casas:2001sr}. The constrained light neutrino parameters $\delta$, $\theta_{12}$, $\theta_{13}$ and $\theta_{23}$ are fixed at their best-fit value from recent global fit data~\cite{Esteban:2020cvm} where we assume normally ordered light neutrino masses and for simplicity assume the lightest neutrino mass $m_1= 0\,\text{eV}$. While we fix the Majorana phases to be $50^\circ$ and $120^\circ$, they do not significantly affect the resulting baryon asymmetry. For the right-handed neutrino masses, we choose an intermediate-mass scale $m_{N_1}\sim 10^{7}\,\text{GeV}$, commensurate with the standard case studied in Ref.~\cite{Moffat:2018wke}. We fix $m_{N_2}$ and $m_{N_3}$ to reduce the Higgs fine-tuning measure while ensuring that leptogenesis occurs well beyond the resonant regime (the decay widths of $N_1$, $N_2$ and $N_3$ are approximately $10^{-5}\,\text{GeV}$, $10^{-3}\,\text{GeV}$ and $10^{-2}\,\text{GeV}$, respectively, which are much smaller than the $N_1$--$N_2$ mass splitting of approximately $10^5\,\text{GeV}$). The remaining Casas-Ibarra parameters, $x_i$ and $y_i$, are chosen to provide a reduced fine-tuning to the light neutrino masses.  Finally, the benchmark has an initial ratio of temperatures of $\kappa = T_{N_1} / T_\text{SM} = 10$, which gives approximately the maximum achievable baryon asymmetry.  We note that if we assumed standard thermal leptogenesis, in the absence of a hot sector, this benchmark point significantly under-produces the baryon asymmetry. We now numerically integrate the evolution equations, show the evolution of various quantities for this benchmark point and investigate the impact of varying one or two parameters at a time on the final baryon asymmetry. 
\subsection{$N_1$ in Kinetic Equilibrium Only}
\label{sec:results-KE-only}
%
We first focus on the case where $N_1$ is only in kinetic equilibrium with itself.  Note that the scenario differs from that considered by Ref.~\cite{Bernal:2017zvx}, where $N_1$ is in equilibrium. 
\begin{figure}[t]
  \centering
  \includegraphics[width=.8\textwidth]{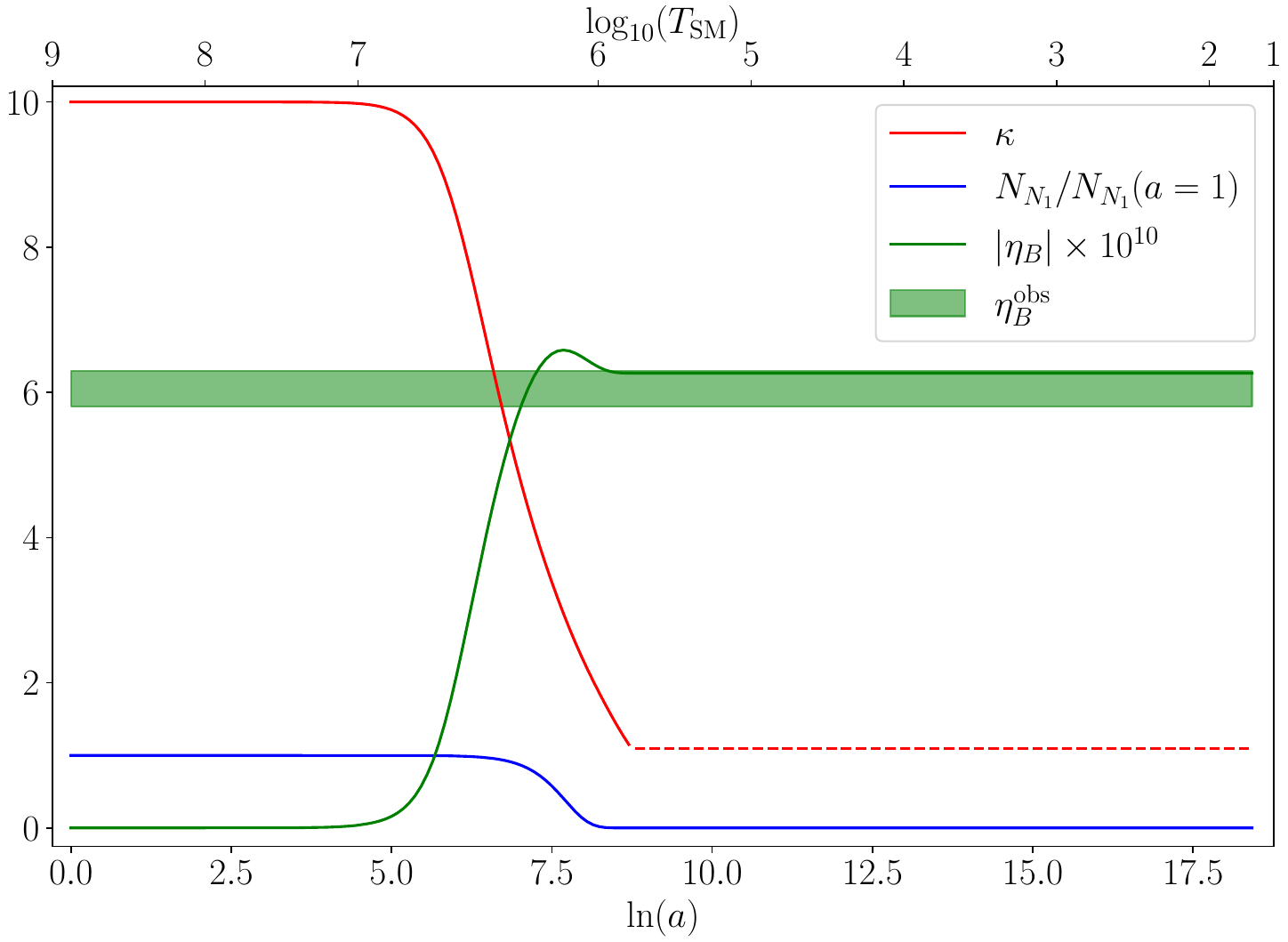}
  \caption{
  Evolution of $|\eta_B|$, $N_{N_1}$ and $\kappa$ for initial $\kappa_\text{in} = 10$ when $N_1$ is only in kinetic equilibrium in the hot sector, for our benchmark point. When the number density approaches zero, $\kappa$ is set to $1$ (dashed line). The green band indicates the baryon-to-photon ratio at the $3\sigma$ level. }
  \label{fig:LeptoEvol}
\end{figure}

In \cref{fig:LeptoEvol} we show the evolution of the temperature ratio $\kappa$ (red), the number density of the lightest right-handed neutrino $N_{N_1}$ (blue), and the baryon asymmetry $\eta_B$ (green) as a function of $\ln a$ (where we set $a=1$ at the beginning of our simulation) for the benchmark point in \cref{tab:params}.  We also show the corresponding SM temperature on the top axis. 
 We see that at early times before $N_1$ has started to decay, at $T_\text{SM}\gtrsim 10^7\,\text{GeV}$ and $T_{N_1} \gtrsim 10^8\,\text{GeV}$, the temperature ratio, the $N_1$ comoving number density and the baryon asymmetry remain constant.  When the $N_1$ population starts to decay, at $T_\text{SM} \sim 10^7\,\text{GeV}$ and $T_{N_1} \sim 10^8\,\text{GeV}$, the decay starts to put the two sectors into kinetic equilibrium and the temperature ratio $\kappa$ begins to fall.  The baryon asymmetry immediately starts rising and overshoots the observed value around halfway through the decay.  It then reduces slightly due to the effect of the washout terms.  The temperature ratio approaches $\kappa=1$ as the hot sector is depleted, and at some point the $N_1$ abundance goes below the numerical accuracy of our computation.  At this point, we set $\kappa = 1$ (dashed red line) so the system can evolve while avoiding numerical errors. 
 This procedure does not affect the final baryon asymmetry $\eta_B$ as almost all $N_1$ particles have decayed by this time.  
 Note, however, that even though the two sectors are technically in kinetic equilibrium, the hot sector is essentially empty since almost all $N_1$ particles have decayed. 
 We see that our benchmark point produces a baryon asymmetry within the observed $3\,\sigma$ band.  For these parameters we find $\Delta_\eta \approx 855\% $ (indicating that the tree-level mass is $\mathcal{O}(10)$ times larger than the loop level mass) and $\Delta_H \approx 10.4\%$ (indicating that the loop level contribution is $\sim$ TeV) so there is no fine-tuning in the light neutrino masses and very mild fine-tuning in the Higgs mass.
\begin{figure}
  \centering
  \includegraphics[width=\textwidth]{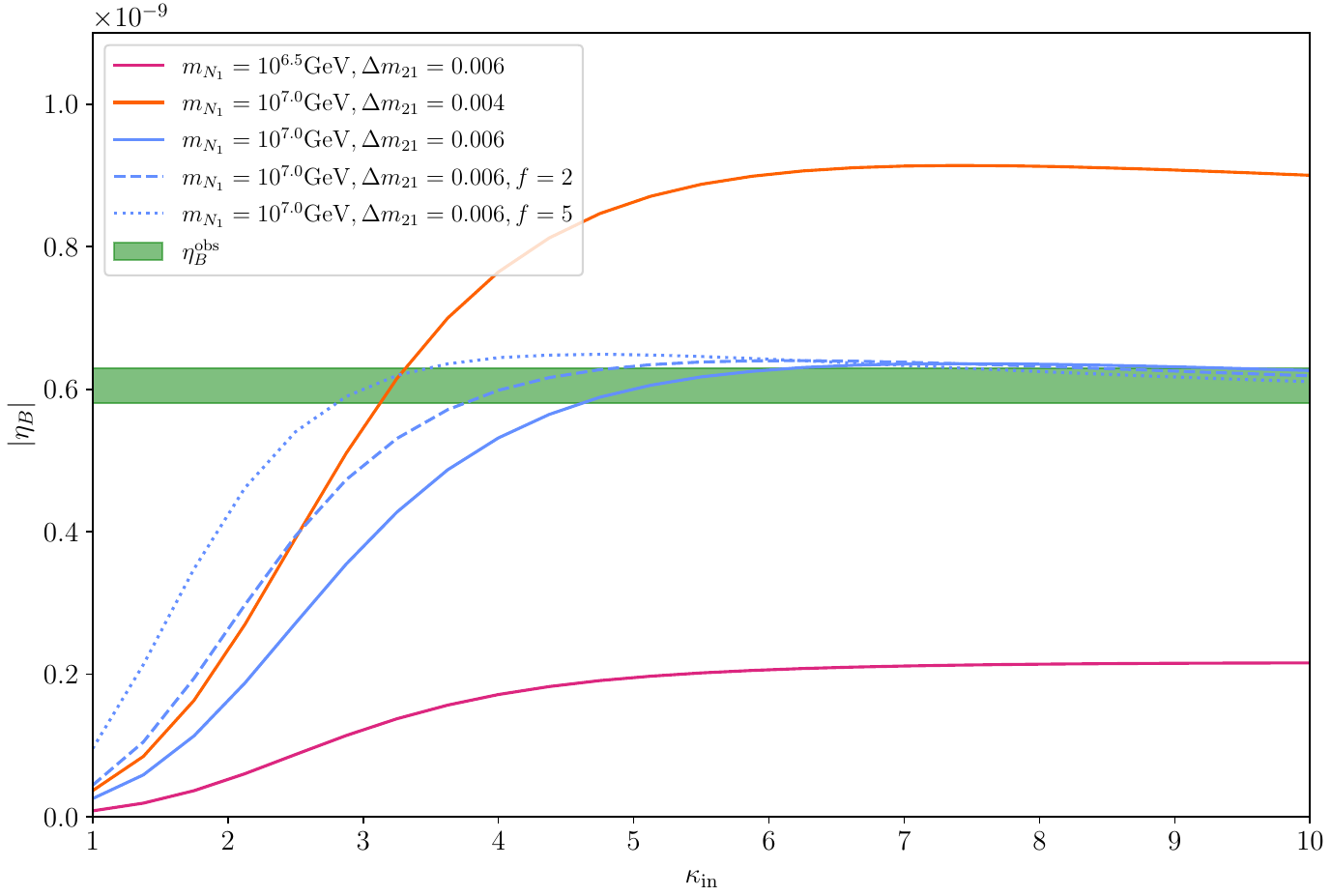}
  \caption{
  The final baryon asymmetry $|\eta_B|$ as a function of the  initial temperature ratio $\kappa_\text{in} = T_{N_{1}}/T_{\rm SM}$ at $a=1$.  The blue line indicates the benchmark point. In the burgundy and orange curves we vary the RHN mass scale and splittings (see text for details), while the dashed (dotted) blue curves indicate non-equilibrium initial abundances of $N_1$, $f \equiv \left(n_{N_1}/n_{N_1}^\text{eq}\right)_\text{in}$.
  }
  \label{fig:KappaScan}
\end{figure}

In \cref{fig:KappaScan} (blue curve) we show the final baryon asymmetry as a function of the initial temperature ratio $\kappa$ for our benchmark point.  
We see that $|\eta_B|$ is far below the observed asymmetry for $\kappa \sim 1$ and first increases with $\kappa$.  The rate of increase reaches a maximum of around $\kappa = 2.7$, where the energy densities of the SM and $N_1$ are approximately equal.  After this point, the baryon asymmetry begins to level off and reaches a maximum around $\kappa \sim 7$. For $1 \lesssim \kappa \lesssim 7$  the initial number density of $N_1$ in the hot sector is larger than for $\kappa = 1$, which enhances the final asymmetry.  For $2.7 < \kappa$ the hot sector dominates the energy density of the universe, and so the Hubble expansion rate (\cref{eq:hubble-toy-model}), which counteracts the increased asymmetry.  Essentially, the energy dump from the hot sector dilutes the baryon asymmetry.  In \cref{fig:KappaScan} we also show the impact of varying important parameters.  
We see that reducing the RHN masses by setting $m_{N_1} = 10^{6.5}\,\text{GeV}$ and keeping the mass splittings fixed  (burgundy curve) significantly reduces the generated asymmetry, as is typical in leptogenesis since the Yukawa matrix $Y \propto M_N$. The $N_1$--$N_2$ mass splitting is defined as $\Delta m_{21} = \log_{10}(m_{N_2}/[{\rm GeV}])-\log_{10}(m_{N_1}/[{\rm GeV}])$, with $\Delta m_{31}$ defined analogously and preserved at $0.4$. Reducing the right-handed neutrino masses reduces the fine-tuning to $\Delta_H = 58 \%$, which is a bit better than our benchmark point, but does not reproduce the observed baryon asymmetry.  
We also see that a smaller mass splitting between $N_1$ and $N_2$ (orange curve) enhances the asymmetry. This also very slightly decreases the amount of fine-tuning needed in the Higgs sector to $\Delta_H = 10.4\,\%$, because the $N_2$ state is lighter.  When fitting the light neutrino masses, the smaller $N_1$ -- $N_2$ mass splitting leads to a larger $Y_{\mu 1}$ and $Y_{\mu 3}$ and a smaller $Y_{\mu 2}$ (with the other Yukawas remaining approximately constant). Since the $Y_{\alpha 1}$ couplings have the largest impact on the generated asymmetry, this leads to an overall increase.  If one wanted to find the minimal fine-tuning possible in this scenario, this could potentially be achieved by reducing the mass splitting further (while remaining out of the resonant leptogenesis regime), which boosts the asymmetry, while reducing the overall RHN mass scale, which reduces the asymmetry and would further improve the fine-tuning in the SM Higgs sector. 
In the dashed and dotted blue lines, we show the impact of changing the initial $N_1$ abundance to 2 and 5 times its equilibrium abundance.  We see that as the initial abundance increases, the asymmetry increases faster and levels off around a similar maximum value, but at a lower $\kappa_\text{in}$.  Increasing the initial abundance is in many ways similar to increasing the hot sector temperature, as both lead to a higher initial number density of $N_1$ and an increased energy density in the hot sector.  While the number and energy densities scale differently with temperature, this does not lead to an increased final asymmetry at large $\kappa_\text{in}$.  In fact, for a higher initial abundance, the asymmetry drops slowly at large $\kappa_\text{in}$.

\begin{figure}
  \centering
  \includegraphics[width=\textwidth]{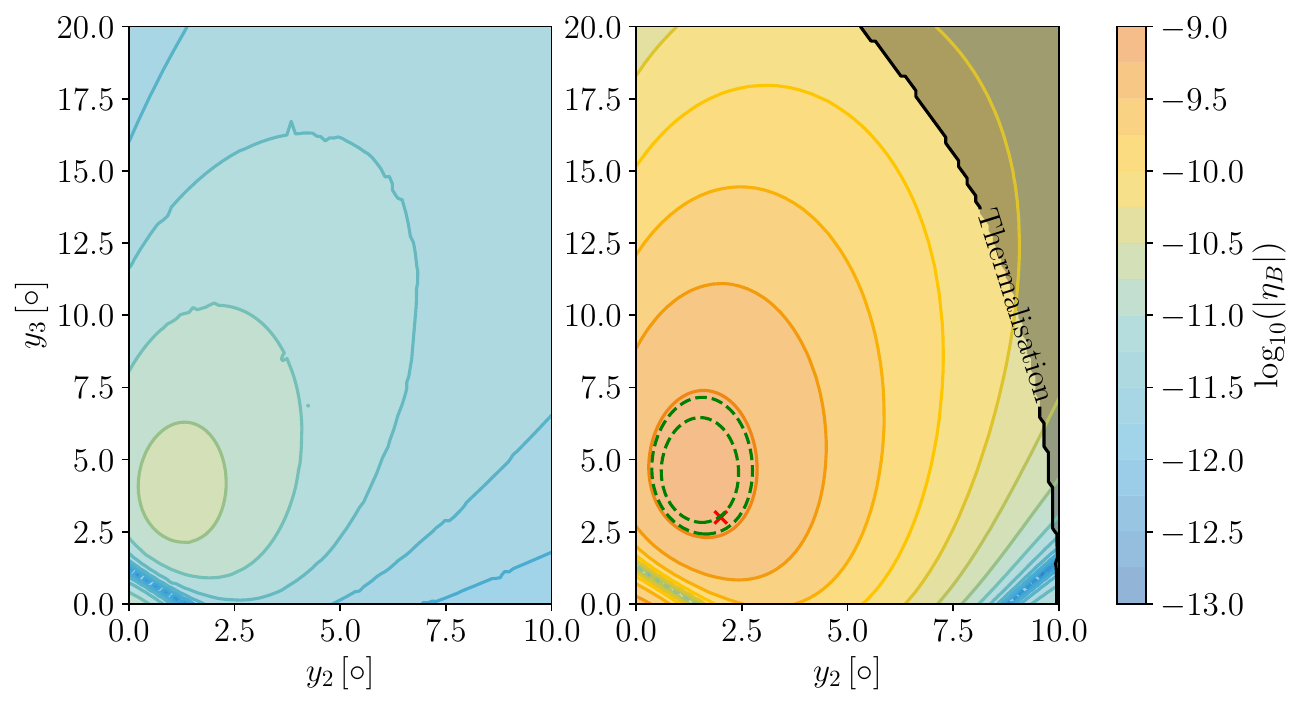}
  \caption{
  Values of $\eta_B$ for standard leptogenesis (left) and hot leptogenesis (right) for $\kappa_\text{in}=10$ produced with the RHN in kinetic equilibrium only. The green dashed contours corresponding to $\eta_B$ produced at $(5.8-6.3)\times 10^{-10}$ \cite{Yeh:2022heq,PDG_23} and the red cross indicates our benchmark point. The greyed-out region represents when the non-thermalisation assumption no longer holds, such that hot leptogenesis may not be viable. $\Delta_H \sim 10.4\,\%$ and $\Delta_\nu \sim 855\,\%$ throughout the plot.}
  \label{fig:KEScan}
\end{figure}

Finally, in \cref{fig:KEScan} we show the baryon asymmetry in $y_2$ and $y_3$, chosen for their impact on the baryon asymmetry. On the left, we show the standard leptogenesis case where $T_{N_1} = T_\text{SM}$ and the $N_1$ are in chemical equilibrium with themselves, while on the right we show the results for hot leptogenesis with an initial $\kappa = 10$ and where $N_1$ particles are only in kinetic equilibrium with themselves.  The benchmark point is indicated by a red cross. In the left panel, we see that in this region of parameter space standard leptogenesis under-produces the baryon asymmetry by more than an order of magnitude. 
In the right panel, we see that $\eta_B$ is enhanced by a factor of $\sim 50$ compared to the standard case and the observed baryon asymmetry can be produced in this parameter space (the dashed green contours give the 3$\sigma$ range). Importantly, this is away from the regime in the top-right where Higgs-mediated $N_1 \ell \to N_1 \ell$ and lepton-mediated $N_1 H \to N_1 H$ elastic scattering processes equilibrate the SM and the hot sector, indicated by the greyed-out `Thermalisation' region. The fine-tuning in both the Higgs mass and the neutrino masses do not depend strongly on the parameters $y_2$ and $y_3$ so remain approximately equal to those of the benchmark point.
\subsection{$N_1$ in Kinetic and Chemical Equilibrium}
We now briefly turn to the scenario where the hot sector is in both kinetic and chemical equilibrium with itself. As described above, in this scenario we assume $n_{N_1} = n_{N_1}^\text{eq}$ throughout and use this relation to find the evolution of the hot sector temperature $T_{N_1}$.
\begin{figure}
  \centering
  \includegraphics[width=\textwidth]{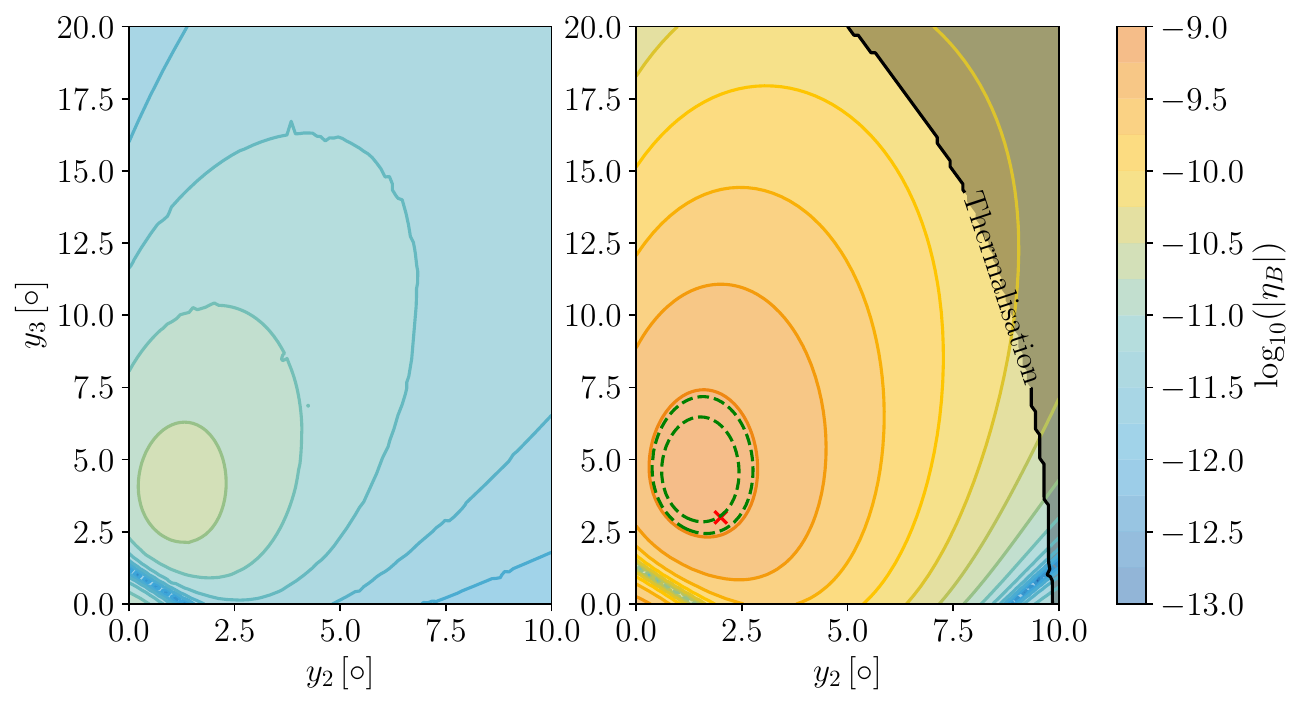}
  \caption{
  Values of $\eta_B$ for standard leptogenesis (left) and hot leptogenesis (right) for $\kappa_\text{in}=10$ produced with the RHN in kinetic and chemical equilibrium. The green dashed contours corresponding to $\eta_B$ produced at $(5.8-6.3)\times 10^{-10}$ \cite{Yeh:2022heq,PDG_23} and the red cross indicates our benchmark point. The greyed out region represents when the non-thermalisation assumption no longer holds, such that hot leptogenesis may not be viable. $\Delta_H \sim 10.4\,\%$ and $\Delta_\nu \sim 855\,\%$ throughout the plot.
  }
  \label{fig:CEScan}
\end{figure}

Even though we are in a different region of parameter space, pink stars in \cref{fig:TEConstraints}, these parameters do not strongly impact the $N_1$ evolution and baryon asymmetry generation, which depend on the parameters in \cref{tab:params}.  Taking the initial $N_1$ abundance to be $N_{N_1}^\text{eq}$ and using the benchmark parameters in \cref{tab:params}, we find that the $\kappa$, $N_{N_1}$ and $\eta_B$ evolution is virtually identical to \cref{fig:LeptoEvol}.  For this reason, we do not show it, but instead just show the results of a parameter scan in $y_2$ and $y_3$ in \cref{fig:CEScan}. We see that the difference with \cref{fig:KEScan} is very small: at most a factor $4$ in the baryon asymmetry near the thermalisation region, but at the percent level in the region where the observed $\eta_B$ is produced.
The thermalisation region has moved slightly to the left, indicating that it is easier for the hot sector to come into kinetic equilibrium with the SM sector in this scenario. 
The fine-tuning, in this case, is identical to that in \cref{sec:results-KE-only}.

Overall, we see that phenomenologically it does not make a significant difference whether only kinetic equilibrium, or both kinetic and chemical equilibrium, are maintained during $N_1$ decay.  Both scenarios can produce the observed baryon asymmetry while avoiding fine-tuning of the neutrino masses or the SM Higgs mass.

\subsection{Further Scenarios}
We now briefly comment on two possible alternative scenarios.  
First, kinetic equilibrium could be established after inflaton decay but not maintained in the hot sector while $N_1$ decays.
This could, for example, occur if a heavy mediator realises fast $2N_1 \to 2 N_1$ scattering shortly after inflaton decay, but is too heavy to maintain it at $T_{N_1} \sim m_{N_1}$. In this scenario, one in principle has to compute the evolution for the full set of momentum modes.
However, in the absence of additional processes affecting the sector, the assumption of a thermal distribution may be reasonable until the $N_1$ start to decay. As was shown in \cite{Hahn-Woernle:2009jyb}, in vanilla leptogenesis the assumption of kinetic equilibrium when it is not realised underestimates the final baryon asymmetry by a tiny amount, because low momentum $N_1$ particles are more efficiently produced than accounted for in a thermal distribution. In our present scenario, we may expect that large momentum $N_1$ particles decay earlier, and thus the produced asymmetry is overestimated by a small amount if kinetic equilibrium is assumed. 

Second, there is the possibility of $N_2$ and $N_3$ starting with temperature at or around $T_{N_1}$ and with an approximately equilibrium abundance, but where all three particles are weakly coupled and do not establish kinetic equilibrium with the SM.  While Ref.~\cite{Bernal:2017zvx} shows that the observed light neutrino masses mean that, in a type-I see-saw scenario, the decay rate of $N_2$ and $N_3$ must be larger than Hubble at temperatures around their masses, we find that they do not have enough time to fully equilibrate before decay. It is therefore possible that this scenario may lead to an asymmetry which is larger by up to a factor of three compared to the results we find here, further reducing the required fine-tuning.

\section{Conclusions}
\label{sec:conclusions}
In this work, we have studied a class of leptogenesis scenarios in which the sector containing the lightest right-handed neutrino ($N_1$) establishes kinetic equilibrium with itself at a temperature higher than that of the Standard Model (SM) sector. We have motivated this setup by considering the decay of the inflaton, which can lead to two sectors with similar but distinct temperatures. Higher temperatures in the $N_1$ sector enhance the number density of $N_1$ particles and can lead to an enhanced baryon asymmetry. With this setup, the observed baryon asymmetry can be generated without the significant fine-tuning of the light neutrino masses and the SM Higgs boson mass present in standard leptogenesis. We have checked that inflaton-mediated energy exchange between the sectors is not fast enough to equilibrate them after inflation. 

In \cref{sec:model-of-hot-leptogenesis} we described a toy model that can realise such a scenario of hot leptogenesis, introducing a new scalar field $\phi$ responsible for mediating self-interactions of $N_1$ and maintaining its kinetic equilibrium. We explore two regimes: one where the hot sector containing $N_1$ is only in kinetic equilibrium with itself and another where it is in both kinetic and chemical equilibrium. We derived the relevant evolution equations to track the relevant quantities and compute the resulting baryon asymmetry in both of these scenarios.

Our numerical analysis reveals that both scenarios can produce the observed baryon asymmetry with minimal fine-tuning. As expected, the enhancement of the baryon asymmetry can be up to a factor of around 50, and the observed asymmetry can then be achieved for smaller right-handed neutrino masses and couplings. These scenarios therefore reduce the fine-tuning required in both the Higgs and neutrino sectors compared to the standard leptogenesis scenario. This confirms that non-resonant leptogenesis is viable and efficient in producing the observed baryon asymmetry under our model assumptions.

Comparing the numerical results in the two cases, we find that the results for kinetic only and kinetic and chemical equilibrium are similar. This can be understood from the fact that we assume $n_{N_1} = n_{N_1}^\text{eq}$ as an initial condition for the former case, which is preserved until $T \sim m_{N_1}$, right before the decays happen. Thus, we expect that chemical equilibrium can be a reasonable approximation in the case where it is not realised or maintained. 

We finally note that our computations are also likely to be a good approximation to the case where  $N_1$ are not in kinetic equilibrium with themselves when they decay and that it may be possible that $N_2$ and $N_3$ are present in the hot sector and could lead to an enhanced baryon asymmetry.

\acknowledgments

AB, DC and JT are supported by the STFC under Grant No.~ST/T001011/1. MJB and DC acknowledge support from Royal Society International Exchanges grant IESR1221078.  MJB also acknowledges support from the University of Massachusetts Amherst, the DIVA Program at the IPPP Durham University, and is grateful to the IPPP for their hospitality during the completion of this work.

\appendix

\section{Cross-Sections, Decay Rates and Thermal Averaging}
\label{sec:cross-sections-thermal-averaging}

In this appendix, we give details of the cross-sections, decay rates and thermal averaging we use in our computations.

\subsection{Cross-Sections and Decay Rates}\label{App1}

In \cref{sec:model-of-hot-leptogenesis} we examine a number of processes that contribute to energy exchange (via elastic scattering) or particle number exchange (via number-changing processes). 
 Here, we give the relevant cross-sections which we calculated with the help of \texttt{FeynCalc}~\cite{Shtabovenko:2016sxi}.

The Feynman diagrams for the elastic scattering processes $N_1 N_{2,3} \to N_1 N_{2,3}$ are given in \cref{fig:feyn-NNi-kinetic-eq} (a) and (b), and the cross-section is
\begin{multline}
    \label{eq:cs-N1NitoN1Ni}
    \sigma_{N_1N_{2,3} \to N_1 N_{2,3}} (s)= \frac{(y_\phi^1)^2 (y_\phi^{2,3})^2}{4 \pi  m_\phi^2 s^2 \left(m_\phi^2+s\right)} \Bigg[s \left(4 m_{N_{1}}^2 \left(4 m_{N_{2,3}}^2-m_\phi^2\right)+m_\phi^2 \left(2 m_\phi^2-4 m_{N_{2,3}}^2+s\right)\right)\\
    +2 m_\phi^2 \left(m_\phi^2+s\right) \left(m_\phi^2-2 \left(m_{N_{1}}^2+m_{N_{2,3}}^2\right)\right) \log \left(\frac{m_\phi^2}{m_\phi^2+s}\right)\Bigg]
    \,,
\end{multline}
where $s$ is the squared centre-of-mass energy.  This process is relevant for considering whether the SM and hot sectors come into equilibrium via elastic scattering.

\Cref{fig:feyn-NNi-kinetic-eq} (a), (b) and (c) contribute to the elastic scattering process $N_1 N_1 \to N_1 N_1$, leading to the cross-section
\begin{multline}
    \label{eq:cs-2N1to2Ni}
  \sigma_{2N_1 \to 2N_1}(s) = \frac{(y_\phi^1)^4}{16 \pi s^2 m_\phi^2 (m_\phi^2-s)^2 (m_\phi^2+s) (2 m_\phi^2+s)} \Bigg[m_\phi^2(m_\phi^4-s^2)(24 m_{N_{1}}^4(s-2m\phi^2)\\
  +16 m_{N_{1}}^2(5s^2+2m_\phi^2 s - 4m_\phi^4) + 16 m_\phi^6 -8m_\phi^4 s - 5m_\phi^2 s^2) \log{\left(\frac{m_\phi^2}{m_\phi^2 + s} \right)}\\
  +s(2m_\phi^2+s)(16 m_{N_{1}}^4 (6 m_\phi^4 - 9 m_\phi^2 s + 5s^2) - 8 m_{N_{1}}^2 (4 m_\phi^6 - 9 m_\phi^4 s + 7 m_\phi^2 s^2 )\\ +8m_\phi^8 - 12 m_\phi^6 s + 3 m_\phi^4 s^2 + 3m_\phi^2 s^3 )\Bigg] \, .
\end{multline}
This process can maintain kinetic equilibrium in the hot sector.

Now we will consider number-changing processes that can maintain chemical equilibrium. 
 The decay rate for $\phi \to 2 N_1$ shown in \cref{fig:feyn-NN-chemical-eq} (a) in the $\phi$ rest frame is given by
\begin{align}
    \label{eq:gamma-2N1tophi}
    \Gamma_{\phi \to 2 N_1}^0
    =
    \frac{(y_\phi^1)^2
    (m_\phi^2-4 m_{N_1}^2)^{\frac{3}{2}}}{8 \pi m_\phi^2}
    \,.
\end{align}
The cross-section for $2N_1\to 2\phi$ shown in \cref{fig:feyn-NN-chemical-eq} (b) and (c) is 
\begin{multline}
    \label{eq:cs-2N1to2phi}
  \sigma_{2N_1 \to 2\phi}(s) = \frac{y_\phi^4 }{64 \pi s \left(s-4 m_{N_{1}}^2\right)}\Bigg[-\frac{ 2\sqrt{s-4 m_{N_{1}}^2} \left(16 m_{N_{1}}^4+2 m_{N_{1}}^2 \left(s-8 m_\phi^2\right)+3 m_\phi^4\right)}{m_{N_{1}}^2 \sqrt{s-4 m_\phi^2}+m_\phi^4}\\
  +\frac{\left(s^2 -4 m_\phi^2 s + 6 m_\phi^4 -32 m_{N_{1}}^4+16 m_{N_{1}}^2 \left(s-m_\phi^2\right)\right)}{\left(s-2 m_\phi^2\right) } \log \left(\frac{\sqrt{\left(s-4 m_{N_{1}}^2\right) \left(s-4 m_\phi^2\right)}-2 m_\phi^2+s}{\sqrt{\left(s-4 m_{N_{1}}^2\right) \left(s-4 m_\phi^2\right)}+2 m_\phi^2-s}\right)\Bigg]\,.
\end{multline}
The cross-section for the $s$-channel process $2N_1 \to 3 \phi$ shown in \cref{fig:feyn-NN-chemical-eq} (d) is \cite{Romao}
\begin{equation}
    \label{eq:NNphi3}
  \begin{split}
    \sigma_{2N_1 \to 3\phi}(s) =& \frac{\lambda ^2 m_{N_{1}}^2 y_\phi^2 }{6144 \pi^2 s^2 (m_\phi-\sqrt{s})^2(m_\phi+\sqrt{s})^{3/2}} 
    \sqrt{\frac{s(3 m_\phi-\sqrt{s})}{s-4 m_{N_{1}}^2}}
    \\ &
    \Bigg[(m_\phi+\sqrt{s})(3 m_\phi^2+s) \tilde{E}(m_\phi,s)
  -(m_\phi-\sqrt{s})^2)(3 m_\phi+\sqrt{s})\tilde{F}(m_\phi,s)\Bigg]\,,
  \end{split}
\end{equation}
where $\tilde{E}(m_\phi,s)$ is defined as
\begin{equation}
\tilde{E}(m_\phi,s) = E\left(\arcsin{\left(\frac{1}{4} \sqrt{\frac{\left(3 m_\phi-\sqrt{s}\right) \left(m_\phi+\sqrt{s}\right)^3}{m_\phi^3 \sqrt{s}}}\right)}, \frac{16 m_\phi^3 \sqrt{s}}{\left(3 m_\phi-\sqrt{s}\right) \left(m_\phi+\sqrt{s}\right)^3}\right)
\,,
\end{equation}
and $E(x, y)$ is the incomplete elliptic integral of the second kind, and similarly for $\tilde{F}(m_\phi,s)$ where $F(x, y)$ is the incomplete elliptic integral of the first kind and the arguments are related in the same way.

\subsection{Thermal Averaging}\label{thermalav}

For the rates considered in \cref{sec:kinetic-and-chemical-equilibria} we need the thermally averaged cross-sections for the initial states $2N_1$ and $N_1 N_{2,3}$, where $N_1$ and $N_{2,3}$ are at different temperatures.

The thermal averaged cross-section for two identical incoming $N_1$ particles is given in Ref.~\cite{Gondolo:1990dk},
\begin{equation}
    \label{eq:average}
  \langle \sigma v \rangle = \frac{1}{8 m_{N_{1}}^4  T_{N_1} K_2^2\left(\frac{m_{N_{1}}}{T_{N_1}}\right)} \int_{4m_{N_{1}}^2}^\infty ds \, \sigma(s)(s-4m_{N_{1}}^2)\sqrt{s} K_1\left(\frac{\sqrt{s}}{T_{N_1}}\right) \, ,
\end{equation}
where $K_1$ and $K_2$ are modified Bessel functions of the first and second kind, respectively. 

For two incoming particles of different masses and different temperatures, we generalise the results in Refs.~\cite{Gondolo:1990dk,Edsjo:1997,Cannoni:2013}.  For the case of the initial state $N_1 N_2$, the thermal average is given by
\begin{equation}
\langle \sigma \bar{v}\rangle
= 
\frac{\int \sigma \bar{v} f_{N_1} f_{N_2} d^3\mathbf{p}_{N_1} d^3\mathbf{p}_{N_2}}{\int f_{N_1} f_{N_2} d^3\mathbf{p}_{N_1} d^3\mathbf{p}_{N_2}}
\end{equation}
where $\bar{v}$ is the Møller velocity.  We then neglect quantum statistics and take the approximation that $f_{N_1}$ and $f_{N_2}$ are given by Maxwell-Boltzmann distributions.  This assumption lets us perform the integrals analytically and  introduces minimal errors in our results.  We first compute the denominator,
\begin{align}
        \int f_{N_1} f_{N_2} d^3\mathbf{p}_{N_1} d^3\mathbf{p}_{N_2} 
        &= 16 \pi^2 T_{N_1} T_{\rm SM} m_{N_1}^2 m_{N_2}^2 K_2\left(\frac{m_{N_1}}{T_{N_{1}}}\right) K_2\left(\frac{m_{N_2}}{T_\text{SM}}\right)
        \,,
\end{align}
where we have used the fact that $E\,dE = |\mathbf{p}|\,d|\mathbf{p}|$ to rewrite the integral.  For the numerator we follow the computation in \cite{Cheek_2022}. 
We change coordinates from $E_{N_1}$ and $E_{N_2}$ to $x_\pm = \frac{E_N}{T_{N_{1}}} \pm \frac{E_{N_2}}{T_\text{SM}}$, where the upper limit of the $x_-$ integration is
\begin{equation}
    x_-^\text{max} 
    = 
    \frac{m_{N_1}^2 T_\text{SM}^2 x_+ +\sqrt{(m_{N_1}^2-s)^2 T_\text{SM} \left( m_{N_1}^2 (T_{N_1}-T_\text{SM}) + T_\text{SM} (T_{N_1} T_\text{SM} x_+^2 - s)\right)}}{m_{N_1}^2 T_\text{SM} (T_\text{SM} - T_{N_{1}}) + s T_{N_{1}} } \, .
\end{equation}
Integration over $x_-$ and $x_+$ then leads to
\begin{equation}\label{eq:thermal-average-N1NitoN1Ni}
    \langle \sigma \bar{v}\rangle = D \int_{(m_{N_1} + m_{N_2})^2}^\infty ds \, \sigma(s) \frac{C}{B} \left(A(1+z)e^{-z} + C\sqrt{B} K_1(z)\right)
\end{equation}
where 
\begin{align}
A &= m_{N_1}^2 T_\text{SM}^2 - m_{N_2}^2 T_{N_1}^2
\\
B &= m_{N_2}^2 T_{N_{1}} (T_{N_{1}} - T_\text{SM}) + m_{N_1}^2 T_\text{SM} (T_\text{SM} - T_{N_{1}}) + s T_{N_{1}} T_\text{SM}
\\
C &= \sqrt{(s-(m_{N_1}-m_{N_2})^2)(s-(m_{N_1}+m_{N_2})^2)} \\
D^{-1} &= 8 m_{N_1}^2 m_{N_2}^2 K_2\left(\frac{m_{N_1}}{T_{N_{1}}}\right) K_2\left(\frac{m_{N_2}}{T_\text{SM}}\right) \, ,
\\
z &= \frac{\sqrt{B}}{T_{N_{1}} T_\text{SM}}
\,.
\end{align}
The thermally averaged cross-section for initial states $N_1 N_3$ is given by replacing $N_2$ with $N_3$.

\section{Fine-Tuning}
\label{app:fine-tuning}

To quantify the degree of fine-tuning present in the neutrino sector, we will adopt a fine-tuning measure that is the inverse of that used in~\cite{Moffat:2018wke}.  The matrix of physical light neutrino masses $M_\nu$ is 
\begin{align}
  M_\nu = M_\nu^\text{tree} + M_\nu^\text{1-loop}
  \,,
\end{align}
where $M_\nu^\text{tree}$ contains the tree-level Lagrangian neutrino masses and $M_\nu^\text{1-loop}$ is the one loop contribution (which is always negative)~\cite{Moffat:2018wke}. The fine-tuning can be measured using, 
\begin{align}
  \Delta_\nu = \frac{
  \sum_{i=1}^3 \text{SVD}[M_\nu]_i
  }{
  \sum_{i=1}^3 \text{SVD}[M_\nu^\text{1-loop}]_i
  }
  \,,
\end{align}
where SVD is the Singular Value Decomposition of the matrix, i.e., $\text{SVD}[M]_i$ is the square root of the $i$-th (real and positive) eigenvalue of $M^\ast M$. If the eigenvalues of $M$ are real and positive, then the singular value decomposition of $M$ simply gives the eigenvalues. Fine-tuning of $\Delta_\nu \approx 1\,\%$ corresponds to, e.g., $M_\nu^\text{tree}  \approx 100 M_\nu$ and $M_\nu^\text{1-loop} \approx 100 M_\nu$, so the tree and loop level masses would cancel to one part in 100.

Analogously for the Higgs sector, we will have 
\begin{align}
  \mu_H^2 \approx (\mu_H^\text{tree})^2 - |\delta \mu^2|
  \,,
\end{align}
where $\mu_H = \frac{m_h}{\sqrt{2}} = 88\,\text{GeV}$ is the effective Higgs mass parameter, $\mu_H^\text{tree}$ is the Lagrangian Higgs parameter and $\delta \mu^2$ is the one-loop correction to the Higgs mass parameter~\cite{Clarke:2015gwa},
\begin{align}
  |\delta \mu^2| 
  \approx
  \frac{1}{4\pi^2}
  \text{Tr}
  \left[
  Y 
  M_N^2
  Y^\dagger
  \right]
  \,.
\end{align}
The degree of fine-tuning in the mass parameters (not the mass squared parameters) can be measured with
\begin{align}
  \Delta_H 
  =
  \sqrt{
  \frac{(\mu_H^\text{tree})^2 - |\delta \mu^2|}{\frac{1}{2}((\mu_H^\text{tree})^2 + |\delta \mu^2|)}
  }
  \approx
  \sqrt{\frac{\mu_H^2}{|\delta \mu^2|}}
  \,,
\end{align}
where we have assumed $\mu_H^2 \ll (\mu_H^\text{tree})^2, |\delta \mu^2|$. Fine-tuning of $\Delta_H = 10\,\%$ corresponds to $|\delta \mu| = 10 \mu_H = 880\,\text{GeV}$.

\bibliographystyle{JHEP}
\bibliography{refs}

\end{document}